\documentclass{aa} 
\usepackage{geometry}
\usepackage{natbib}
\graphicspath{{EPS/}}
\usepackage{psfig}
\usepackage{epsfig}
\usepackage{abbrevjournaux}
\usepackage{aas_macros}
\usepackage{theorem,algorithm,algorithmic}
\usepackage{caption}
\usepackage{amsmath}
\usepackage{bm}

\usepackage{siunitx}
\usepackage{xspace}

\newcommand*{\Euclid}{\textit{Euclid}\xspace}

\newcommand{\simpropinn}[2]{\mathrel{\vcenter{
  \offinterlineskip\halign{\hfil$##$\cr
    #1\propto\cr\noalign{\kern2pt}#1\sim\cr\noalign{\kern-2pt}}}}}

\renewcommand*\vec[1]{\ensuremath{\boldsymbol{#1}}}

\newcommand*{\bea}{\begin{align}}
\newcommand*{\eea}{\end{align}}
\newcommand*{\be}{\begin{equation}}
\newcommand*{\ee}{\end{equation}}
\newcommand*{\bc}{\begin{center}}
\newcommand*{\ec}{\end{center}}
\newcommand*{\bfig}{\begin{figure}}
\newcommand*{\bfigs}{\begin{figure}\sidecaption}
\newcommand*{\efig}{\end{figure}}
\newcommand*{\btable}{\begin{table}}
\newcommand*{\etable}{\end{table}}
\newcommand*{\bi}{\begin{itemize}}
\newcommand*{\ei}{\end{itemize}}
\newcommand*{\ben}{\begin{enumerate}}
\newcommand*{\een}{\end{enumerate}}
\newcommand*{\bd}{\begin{description}}
\newcommand*{\ed}{\end{description}}

\newcommand*{\AckInstitutions}{support of a number of agencies and
  institutes that have supported the development of \Euclid. A detailed
  complete list is available on the \Euclid\ web site 
(\texttt{http://www.euclid-ec.org}). 
In particular the Academy of Finland, the Agenzia Spaziale Italiana,
the Belgian Science Policy, the Canadian Euclid Consortium, the Centre
National d'Etudes Spatiales, the Deutsches Zentrum f\"ur Luft- and
Raumfahrt, the Danish Space Research Institute, the Funda\c{c}\~{a}o
para a Ci\^{e}nca e a Tecnologia, the Ministerio de Economia y
Competitividad, the National Aeronautics and Space Administration, the
Netherlandse Onderzoekschool Voor Astronomie, the Norvegian Space
Center, the Romanian Space Agency, the State Secretariat for
Education, Research and Innovation (SERI) at the Swiss Space Office
(SSO), and the United Kingdom Space Agency.\xspace}
\newcommand{\AckEC}{The Euclid Consortium acknowledges the European
  Space Agency and the \AckInstitutions}

\begin{document}

\title{\Euclid: Reconstruction of weak-lensing mass maps for non-Gaussianity studies\thanks{This paper is published on behalf of the Euclid Consortium}}

\author{S.~Pires$^{1}$\thanks{\email{sandrine.pires@cea.fr}}\and V.~Vandenbussche$^{1}$\and V.~Kansal$^{1}$\and R.~Bender$^{2\and3}$\and L.~Blot$^{4}$\and D.~Bonino$^{5}$\and A.~Boucaud$^{6}$\and J.~Brinchmann$^{7}$\and V.~Capobianco$^{5}$\and J.~Carretero$^{8}$\and M.~Castellano$^{9}$\and S.~Cavuoti$^{10,11,12}$\and R.~Cl\'edassou$^{13}$\and G.~Congedo$^{14}$\and L.~Conversi$^{15}$\and L.~Corcione$^{5}$\and F.~Dubath$^{16}$\and P.~Fosalba$^{17,18}$\and M.~Frailis$^{19}$\and E.~Franceschi$^{20}$\and M.~Fumana$^{21}$\and F.~Grupp$^{3}$\and F.~Hormuth$^{22}$\and S.~Kermiche$^{23}$\and M.~Knabenhans$^{24}$\and R.~Kohley$^{15}$\and B.~Kubik$^{25}$\and M.~Kunz$^{26}$\and S.~Ligori$^{5}$\and P.B.~Lilje$^{27}$\and I.~Lloro$^{17,18}$\and E.~Maiorano$^{20}$\and O.~Marggraf$^{28}$\and R.~Massey$^{29}$\and G.~Meylan$^{30}$\and C.~Padilla$^{8}$\and S.~Paltani$^{16}$\and F.~Pasian$^{19}$\and M.~Poncet$^{13}$\and D.~Potter$^{24}$\and F.~Raison$^{3}$\and J.~Rhodes$^{31}$\and M.~Roncarelli$^{20,32}$\and R.~Saglia$^{2,3}$\and P.~Schneider$^{28}$\and A.~Secroun$^{23}$\and S.~Serrano$^{17,33}$\and J.~Stadel$^{24}$\and P.~Tallada Cresp\'i$^{34}$\and I.~Tereno$^{35,36}$\and R. Toledo-Moreo$^{37}$\and Y.~Wang$^{38}$}

\authorrunning{S.~Pires et al.}

\institute{{\tiny List of institutions given in Appendix.}}

\abstract{Weak lensing, which is the deflection of light by matter along the line of sight, has proven to be an efficient method for constraining models of structure formation and reveal the nature of dark energy. So far, most weak-lensing studies have focused on the shear field that can be measured directly from the ellipticity of background galaxies. 
However, within the context of forthcoming full-sky weak-lensing surveys such as \Euclid, convergence maps (mass
maps) offer an important advantage over shear fields in terms of cosmological exploitation. While
it carry the same information, the lensing signal is more compressed in the convergence maps than in
the shear field. This simplifies otherwise computationally expensive analyses, for instance, non-Gaussianity studies. 
However, the inversion of the non-local shear field requires accurate control of systematic effects caused by holes in the data field, field borders, shape noise, and the fact that the shear is not a direct observable (reduced shear). We present the two mass-inversion methods that are included in the official Euclid data-processing pipeline: the standard Kaiser \& Squires method (KS), and a new mass-inversion method (KS+) that aims to reduce the information loss during the mass inversion. This new method is based on the KS method and includes corrections for mass-mapping systematic effects. The results of the KS+ method are compared to the original implementation of the KS method in its simplest form, using the Euclid Flagship mock galaxy catalogue. In particular, we estimate the quality of the reconstruction by comparing the two-point correlation functions and third- and fourth-order moments obtained from shear and convergence maps, and we analyse each systematic effect independently and simultaneously.
We show that the KS+ method substantially reduces the errors on the two-point correlation function and moments compared to the KS method. In particular, we show that the errors introduced by the mass inversion on the two-point correlation of the convergence maps are reduced by a factor of about 5, while the errors on the third- and fourth-order moments are reduced by factors of about 2 and 10, respectively.}


\maketitle

\section{Introduction}

Gravitational lensing is the process in which light from background galaxies is deflected as it travels towards us. The deflection is a result of the gravitation of the intervening mass.
Measuring the deformations in a large sample of galaxies offers a direct probe of the matter distribution in the Universe (including dark matter) and can thus be directly compared to theoretical models of structure formation. The statistical properties of the weak-lensing field can be assessed by a statistical analysis of either the shear field or the convergence field.
On the one hand, convergence is a direct tracer of the total matter distribution integrated along the line of sight, and is therefore directly linked with the theory. On the other hand, the shear (or more exactly, the reduced shear) is a direct observable and usually preferred for simplicity reasons.

Accordingly, the most common method for characterising the weak-lensing field distribution is the shear two-point correlation function. It is followed very closely by the mass-aperture two-point correlation functions, which are the result of convolving the shear two-point correlation functions by a compensated filter \citep{2pcf:schneider02} that is able to separate the E and B modes of the two-point correlation functions \citep{wl:crittenden02}.
However, gravitational clustering is a non-linear process, and in particular, the mass distribution is highly non-Gaussian at small scales. For this reason, several estimators of the three-point correlation functions have been proposed, either in the shear field \citep{wl:bernardeau02,wl:benabed06} or using the mass-aperture filter \citep{map:kilbinger05}. The three-point correlation functions are the lowest order statistics to quantify non-Gaussianity in the weak-lensing field and thus provide additional information on structure formation models.

The convergence field can also be used to measure the two- and three-point correlation functions and other higher-order statistics. When we assume that the mass inversion (the 
computation of the convergence map from the measured shear field) is properly conducted, the shear field contains the same information as the convergence maps \citep[e.g.][]{2pcf:schneider02,3pcf:shi11}. While it carries the same information, the lensing signal is more compressed in the convergence maps than in the shear field, which makes it easier to extract and computationally less expensive.
The convergence maps becomes a new tool that might bring additional constraints complementary to those that we can obtain from the shear field.
However, the weak-lensing signal being highly non-Gaussian at small scales, mass-inversion methods using smoothing or de-noising to regularise the problem are not optimal.

Reconstructing convergence maps from weak lensing is a difficult task because of shape noise, irregular sampling, complex survey geometry, and the fact that the shear is not a direct observable.
This is an ill-posed inverse problem and requires regularisation to avoid pollution from spurious B modes.
Several methods have been derived to reconstruct the projected mass distribution from the observed shear field. The first non-parametric mass reconstruction was proposed by \cite{wl:kaiser93} and was further improved by \cite{wl:bartelmann95}, \cite{wl:kaiser95}, \cite{wl:schneider95}, and \cite{wl:squires96}. These linear inversion methods are based on smoothing with a fixed kernel, which acts as a regularisation of the inverse problem. Non-linear reconstruction methods were also proposed using different sets of priors and noise-regularisation techniques \citep{wl:bridle98,wl:seitz98,wl:marshall02,wl:pires09,wl:jullo14, wl:lanusse16}.
Convergence mass maps have been built from many surveys, including
the COSMOS Survey \citep{cosmos:massey07}, 
the Canada France Hawa\"i Telescope Lensing Survey CFHTLenS \citep{cfhtlens:vanwaerbeke13},  the CFHT/MegaCam Stripe-82 Survey \citep{cs82:shan14}, 
the Dark Energy Survey Science Verification DES SV \citep{dessv:chang15,dessv:vikram15,dessv:jeffrey18}, 
the Red Cluster Sequence Lensing Survey RCSLenS \citep{rcslens:hildebrandt16}, and the Hyper SuprimeCam Survey \citep{hsc:oguri18}.
With the exception of \cite{dessv:jeffrey18}, who used the non-linear reconstruction proposed by \cite{wl:lanusse16}, all these methods are based on the standard Kaiser \& Squires method.

In the near future, several wide and deep weak-lensing surveys are planned: Euclid \citep{euclid:laureijs11}, Large Synoptic Survey Telescope LSST \citep{lsst:abell09}, and Wide Field Infrared Survey Telescope WFIRST \citep{wfirst:green12}.
In particular, the Euclid satellite will survey 15 000 deg$^2$ of the sky to map the geometry of the dark Universe.
One of the goals of the Euclid mission is to produce convergence maps for non-Gaussianity studies and constrain cosmological parameters.
To do this, two different mass inversion methods are being included into the official Euclid data processing pipeline. 
The first method is the standard Kaiser \& Squires method (hereafter KS). Although it is well known that the KS method has several shortcomings, it is taken as the reference for cross-checking the results. The second method is a new non-linear mass-inversion method (hereafter KS+) based on the formalism developed in \cite{wl:pires09}. The KS+ method aims at performing the mass inversion with minimum information loss. This is done by performing the mass inversion with no other regularisation than binning while controlling systematic effects.

In this paper, the performance of these two mass-inversion methods is investigated using the Euclid Flagship mock galaxy catalogue (version 1.3.3, Castander F. et al. in prep) with realistic observational effects (i.e. shape noise, missing data, and the reduced shear). The effect of intrinsic alignments is not studied in this paper because we lack simulations that would properly model  intrinsic alignments.
However, intrinsic alignments also need to be considered seriously because they affect second- and higher-order statistics. A contribution of several percent is expected to two-point statistics  \citep[see e.g.][]{ia:joachimi13}.

We compare the results obtained with the KS+ method to those obtained with a version of the KS method in which no smoothing step is performed other than binning. 
We quantify the quality of the reconstruction using two-point correlation functions and moments of the convergence.
Our tests illustrate the efficacy of the different mass-inversion methods in preserving the second-order statistics and higher-order moments.

The paper is organised as follows.
In Sect. 2 we present the weak-lensing mass-inversion problem and the standard KS method.
Section 3 presents the KS+ method we used to correct for the different systematic effects.
In Sect. 4 we explain the method with which we compared these two mass-inversion methods.
In Sect. 5 we use the Euclid Flagship mock galaxy catalogue with realistic observational effects such as shape noise and complex survey geometry and consider the reduced shear to investigate the performance of the two mass-inversion methods. First, we derive simulations including only one issue at a time to test each systematic effect independently. Then we derive realistic simulations that include them all and study the systematic effects simultaneously.
We conclude in Sect. 6.

\section{Weak-lensing mass inversion}
\label{inversion}

\subsection{Weak gravitational lensing formalism}
\label{formalism}
In weak-lensing surveys, the shear field $\gamma({\vec{\theta}})$ is derived from the ellipticities of the background galaxies at position {\vec{\theta}} in the image. The two components of the shear can be written in terms of the lensing potential $\psi({\vec{\theta}})$ as \citep[see e.g.][]{wl:bartelmann01}
\begin{eqnarray}
\label{eq:gamma_psi} 
\gamma_1 & = & \frac{1}{2}\left( \partial_1^2 - \partial_2^2 \right) \psi, \nonumber \\
\gamma_2 & = & \partial_1 \partial_2 \psi,
\end{eqnarray}
where the partial derivatives $\partial_i$ are with respect to the angular coordinates $\theta_i$, $i = 1,2$ representing the two dimensions of sky coordinates. 
The convergence $\kappa({\vec{\theta}})$ can also be
expressed in terms of the lensing potential as
\begin{eqnarray}
\label{eq:kappa_psi}
\kappa =  \frac{1}{2}\left(\partial_1^2 + \partial_2^2 \right) \psi.
\end{eqnarray}
For large-scale structure lensing, assuming a spatially flat Universe, the convergence at a sky position ${\vec{\theta}}$ from sources at comoving distance $r$ is defined by 
\begin{eqnarray}
\label{eq:kappa_r}
\kappa(\vec{\theta}, r) =\frac{3H^2_0\Omega_{\rm m}}{2 c^2}\int_0^r {\rm d}r' \frac{r'(r-r')}{r} \frac{\delta(\vec{\theta}, r')}{a(r')},
\end{eqnarray}
where $H_0$ is the Hubble constant, $\Omega_{\rm m}$ is the matter density, $a$ is the scale factor, and $\delta \equiv (\rho-\bar\rho)/\bar\rho$ is the density contrast (where $\rho$ and $\bar\rho$ are the 3D density and the mean 3D density, respectively).
In practice, the expression for $\kappa$ can be generalised to sources with a distribution in redshift, or equivalently, in comoving distance $f(r)$, yielding
\begin{eqnarray}
\label{eq:kappa}
\kappa(\vec{\theta}) =\frac{3H^2_0\Omega_{\rm m}}{2 c^2}\int_0^{r_{\rm H}} {\rm d}r'p(r')r' \frac{\delta(\vec{\theta}, r')}{a(r')},
\end{eqnarray}
where $r_{\rm H}$ is the comoving distance to the horizon.
The convergence map reconstructed over a region on the sky gives us the integrated mass-density fluctuation weighted by the lensing-weight function $p(r')$,
\begin{eqnarray}
\label{eq:kappa_r}
p(r') =\int_{r'}^{r_{\rm H}} {\rm d}r f(r)\frac{r-r'}{r}.
\end{eqnarray}

\subsection{Kaiser \& Squires method (KS)}  
\label{ks}

\subsubsection{KS mass-inversion problem}

The weak lensing mass inversion problem consists of reconstructing the convergence $\kappa$ from the measured shear field $\gamma$. We can use complex notation to represent the shear field, $\gamma = \gamma_1 + {\rm i} \gamma_2$, and the convergence field, $\kappa = \kappa_{\rm E} + {\rm i} \kappa_{\rm B}$, with $\kappa_{\rm E}$ corresponding to the curl-free component and $\kappa_{\rm B}$ to the gradient-free component of the field, called E and B modes by analogy with the electromagnetic field.
Then, from Eq.~(\ref{eq:gamma_psi}) and Eq.~(\ref{eq:kappa_psi}), we can derive the relation between the shear field $\gamma$  and the convergence field $\kappa$.
For this purpose, we take the Fourier transform of these equations and obtain
\begin{eqnarray}
\label{eq:kappa2gamma}
\hat \gamma = \hat P \, \hat \kappa,
\end{eqnarray}
where the hat symbol denotes Fourier transforms, $\hat P = \hat P_1 + {\rm i} \hat P_2$,
\begin{eqnarray}
\hat{P_1}(\vec{\ell}) & = & \frac{\ell_1^2 - \ell_2^2}{\ell^2}, \nonumber \\
\hat{P_2}(\vec{\ell}) & = & \frac{2 \ell_1 \ell_2}{\ell^2},
\end{eqnarray}
with $\ell^2 \equiv \ell_1^2 + \ell_2^2$ and $\ell_i$ the wave numbers corresponding to the angular coordinates $\theta_i$. 

$\hat P$ is a unitary operator. The inverse operator is its complex conjuguate $\hat P^* = \hat P_1 - {\rm i} \hat P_2$ , as shown by \cite{wl:kaiser93},
\begin{eqnarray}
\label{eq:gamma2kappa}
\hat \kappa =  \hat P^* \, \hat \gamma.
\end{eqnarray}
We note that to recover $\kappa$ from $\gamma,$ there is a degeneracy when $\ell_1 = \ell_2 = 0$. Therefore the mean value of $\kappa$ cannot be recovered if only shear information is available. This is the so-called mass-sheet degeneracy \citep[see e.g.][for a discussion]{wl:bartelmann95}.
In practice, we impose that the mean convergence vanishes across the survey by setting the reconstructed $\ell = 0$ mode to zero. This is a reasonable assumption for large-field reconstruction \citep[e.g.][]{wl:seljak98}.

We can easily derive an estimator of the E-mode and B-mode convergence in the Fourier domain,
\begin{eqnarray}
\label{eq:fourier}
\hat{\tilde \kappa}_{\rm E} &=& \hat P_1 \hat \gamma_1  +  \hat P_2 \hat \gamma_2,\\ \nonumber 
\hat{\tilde \kappa}_{\rm B} &=& - \hat P_2 \hat \gamma_1 +  \hat P_1 \hat \gamma_2.
\end{eqnarray}
Because the weak lensing arises from a scalar potential (the lensing potential $\psi$), it can be shown that weak lensing only produces E modes. However, intrinsic alignments and imperfect corrections of the point spread function (PSF) generally generate both E and B modes. The presence of B modes can thus be used to test for residual systematic effects in current weak-lensing surveys.

\subsubsection{Missing-data problem in weak lensing}
The shear is only sampled at the discrete positions of the galaxies where the ellipticity is measured. 
The first step of the mass map-inversion method therefore is to bin the observed ellipticities of galaxies on a regular pixel grid to create what we refer to as the observed shear maps $\gamma^{\rm{obs}}$.
Some regions remain empty because various masks were applied to the data, such as the masking-out of bright stars or camera CCD defects. In such cases, the shear is set to zero in the original KS method,
\begin{eqnarray}
\label{eq:mask}
\gamma^{\rm{obs}} &=& M \gamma^{\rm n},
\end{eqnarray}
with $M$ the binary mask (i.e. $M = 1$ when we have information at the pixel, $M = 0$ otherwise) and $\gamma^{\rm n}$ the noisy shear maps.
As the shear at any sky position is non-zero in general, this introduces errors in the reconstructed convergence maps.
Some specific methods address this problem by discarding masked pixels at the noise-regularisation step \cite[e.g.][]{cfhtlens:vanwaerbeke13}. However, as explained previously, this intrinsic filtering results in subtantial signal loss at small scales. Instead, inpainting techniques are used in the KS+ method to fill the masked regions (see Appendix A).

\subsubsection{Weak-lensing shape noise}
\label{sect_shape_noise}
The gravitational shear is derived from the ellipticities of the background galaxies.
However, the galaxies are not intrinsically circular, therefore their measured ellipticity is a combination of their intrinsic ellipticity and the gravitational lensing shear. 
The shear is also subject to measurement noise and uncertainties in the PSF correction. All these effects can be modelled as an additive noise, $N = N_1 + {\rm i} N_2$,
\begin{eqnarray}
\gamma^{\rm n} &=& \gamma+ N
\label{eq:noise1}
\end{eqnarray}
The noise terms $N_1$ and $N_2$ are assumed to be Gaussian and uncorrelated with zero mean and standard deviation,
\begin{eqnarray}
\sigma_{\rm n}^i = \frac{\sigma_{\rm \epsilon}}{\sqrt{N_{\rm g}^i}}, 
\label{eq:noise2}
 \end{eqnarray}
where $N_{\rm g}^i$ is the number of galaxies in pixel $i$.
The root-mean-square shear dispersion per galaxy, $\sigma_{\rm \epsilon}$, arises both from the measurement uncertainties and the intrinsic shape dispersion of galaxies. The Gaussian assumption is a reasonable assumption, and $\sigma_{\rm \epsilon}$ is set to 0.3 for each component as is generally found in weak-lensing surveys \citep[e.g.][]{sigmae:leauthaud07, sigmae:schrabback15, sigmae:schrabback18}. The surface density of usable galaxies is expected to be around $n_{\rm g} = 30$ gal. arcmin$^{-2}$  for the Euclid Wide survey \citep{euclid:cropper13}.

The derived convergence map is also subject to an additive noise,
\begin{eqnarray}
\hat{\tilde \kappa}^{\rm n} = \hat P^* \, \hat{ \gamma}^{\rm n} = \hat \kappa + \hat P^* \, \hat{N}.
\label{kappan}
\end{eqnarray}
In particular, the E component of the convergence noise is
\begin{eqnarray}
N_{\rm E}= N_1* P_1 + N_2 * P_2 , 
\label{eq:noise3}
\end{eqnarray}
where the asterisk denotes the convolution operator, and $P_1$ and $P_2$ are the inverse Fourier transforms of $\hat{P_1}$ and $\hat{P_2}$.
When the shear noise terms $N_1$ and $N_2$ are Gaussian, uncorrelated, and with a constant standard deviation across the field, the convergence noise is also Gaussian and uncorrelated.
In practice, the number of galaxies varies slightly across the field. The variances of $N_1$ and $N_2$ might also be slightly different, which can be modelled by different values of $\sigma_{\epsilon}$ for each component. These effects introduce noise correlations in the convergence noise maps, but they were found to remain negligible compared to other effects studied in this paper.

In the KS method, a smoothing by a Gaussian filter is frequently applied to the background ellipticities before mass inversion to regularise the solution.
Although performed in most applications of the KS method, this noise regularisation step is not mandatory. It was introduced to avoid infinite noise and divergence at very small scales.
However, the pixelisation already provides an intrinsic regularisation. This means that there is no need for an additional noise regularisation prior to the inversion. Nonetheless, for specific applications that require denoising in any case, the filtering step can be performed before or after the mass inversion.

\section{Improved Kaiser \& Squires method (KS+)}  
\label{iks}

Systematic effects in mass-inversion techniques must be fully controlled in order to use convergence maps as cosmological probes for future wide-field weak-lensing experiments such as \Euclid. 
We introduce the KS+ method based on the formalism developed in \cite{wl:pires09} and \cite{wl:jullo14}, which integrates the necessary corrections for imperfect and realistic measurements.
We summarise its improvements over KS in this section and evaluate its performance in Sect. \ref{results_1}.

In this paper, the mass-mapping formalism is developed in the plane. 
The mass inversion can also be performed on the sphere, as proposed in \cite{sphere:pichon09} and \cite{sphere:chang18}, and the extension of the KS+ method to the curved sky is being investigated. 
However, the computation time and memory required to process the spherical mass inversion means limitations in terms of convergence maps resolution and/or complexity of the algorithm.
Thus, planar mass inversions remain important for reconstructing convergence maps with a good resolution and probing the non-Gaussian features of the weak-lensing field (e.g. for peak-count studies).

\subsection{Missing data}
\label{KS+}

When the weak-lensing shear field $\gamma$ is sampled on a grid of $N \times N$ pixels, we can describe the complex shear and convergence fields by their respective matrices. In the remaining paper, the notations $\bm \gamma$ and $\bm \kappa$ stand for the matrix quantities.

\begin{figure*}[h]
\vbox{
\centerline{
\hbox{
\psfig{figure=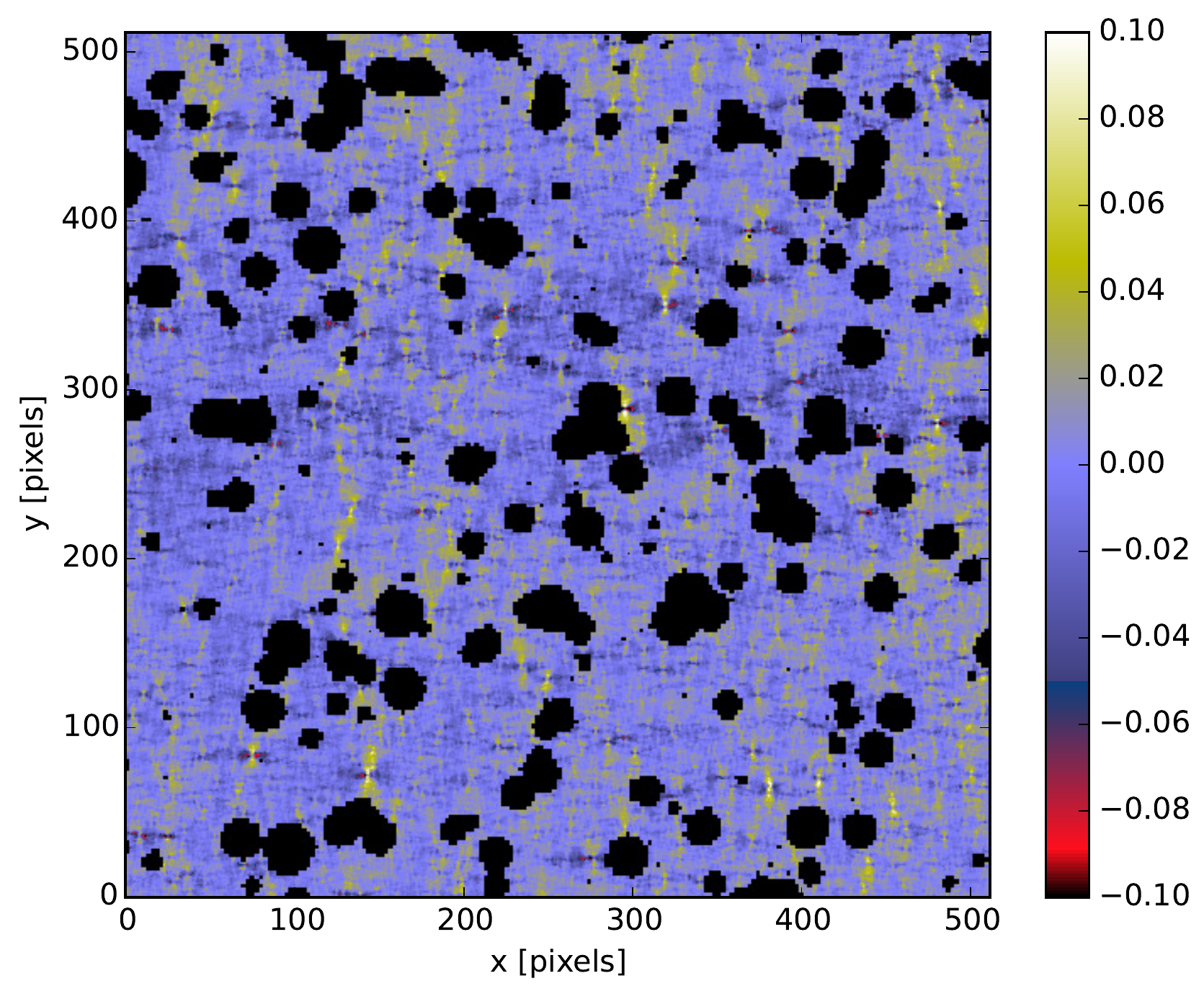,height=6.cm,width=7.cm,clip=}
\hspace{0.2cm}
\psfig{figure=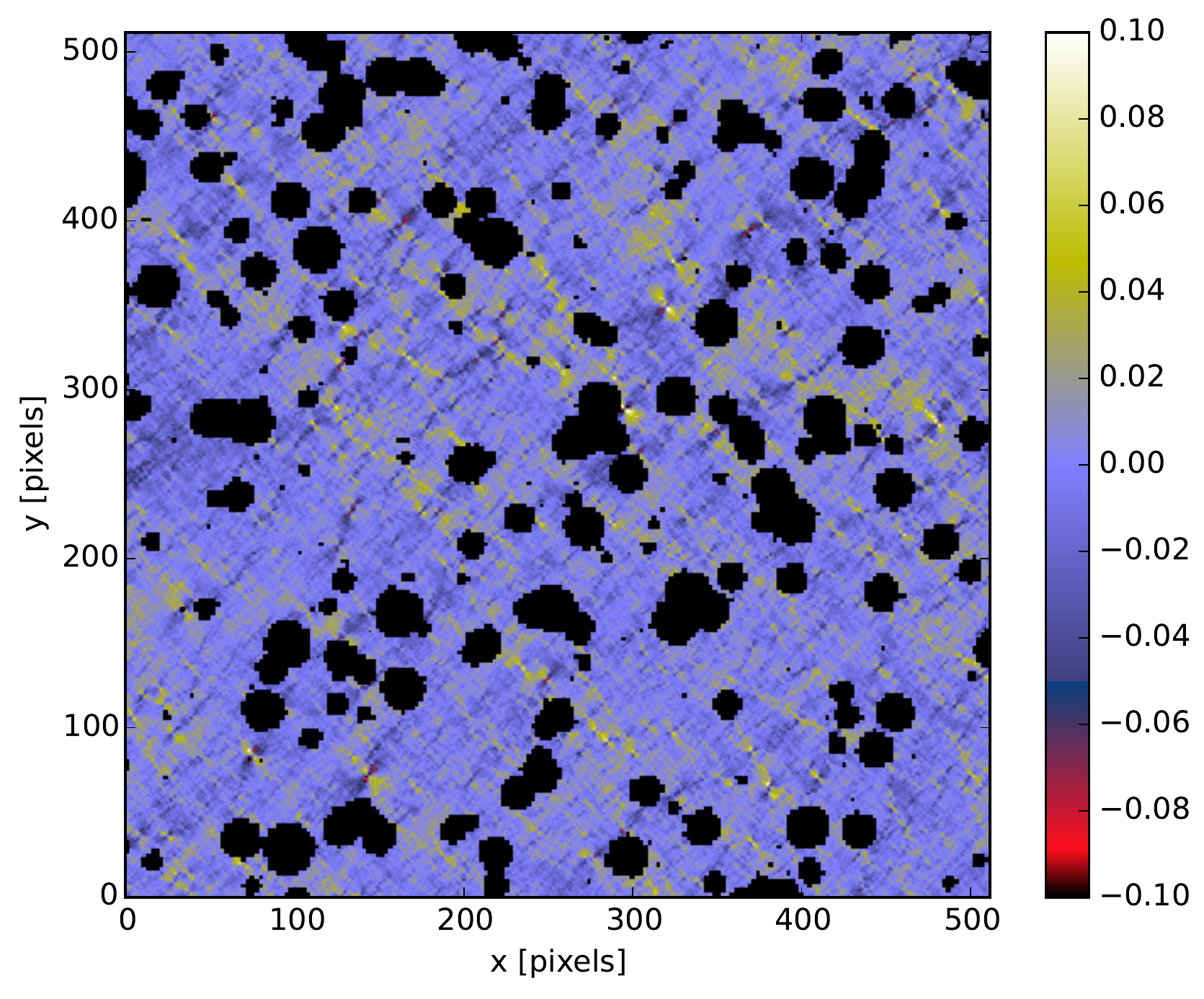,height=6.cm,width=7.cm,clip=}
}}
\vspace{0.3cm}
\centerline{
\hbox{
\psfig{figure=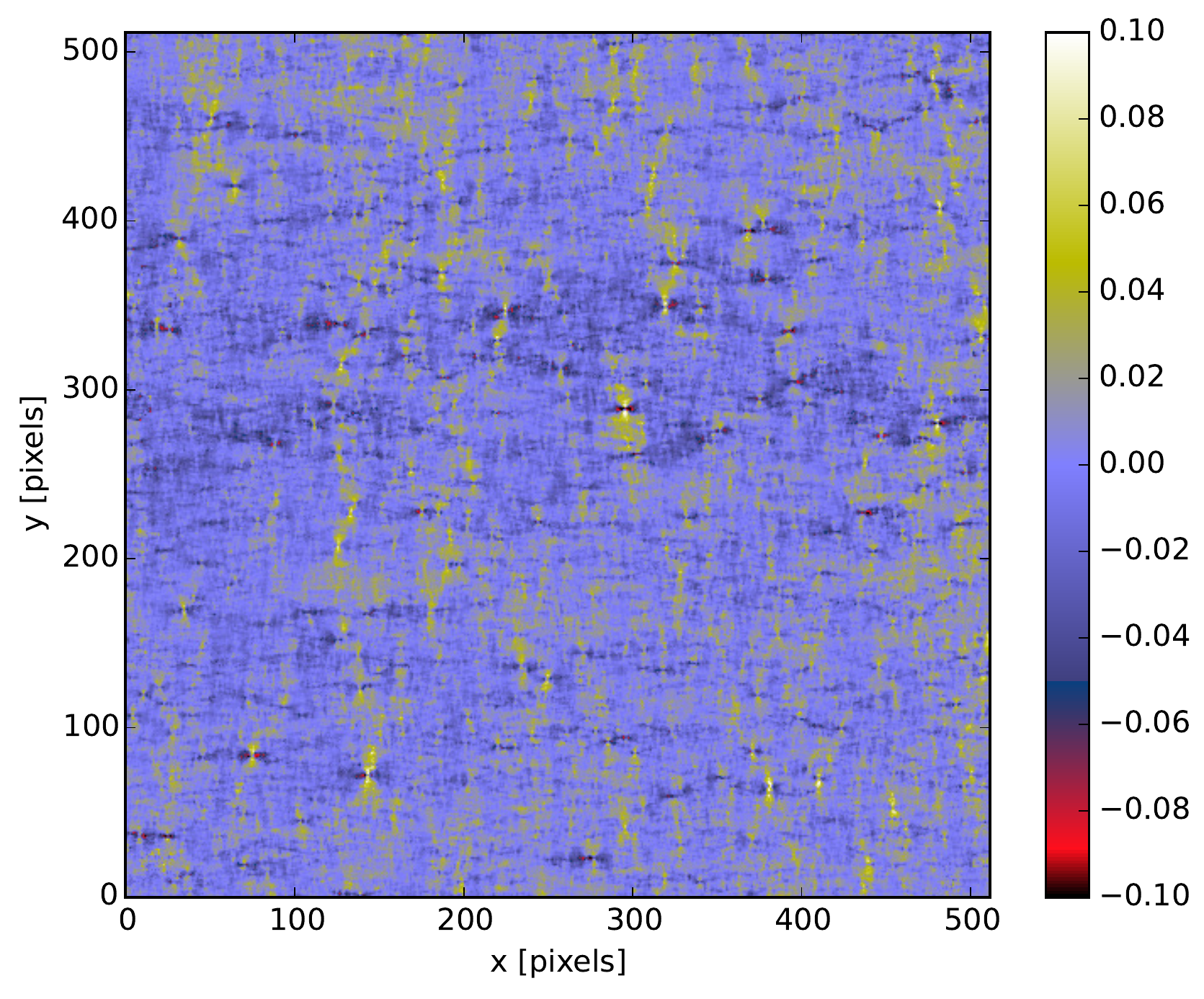,height=6.cm,width=7.cm,clip=}
\hspace{0.2cm}
\psfig{figure=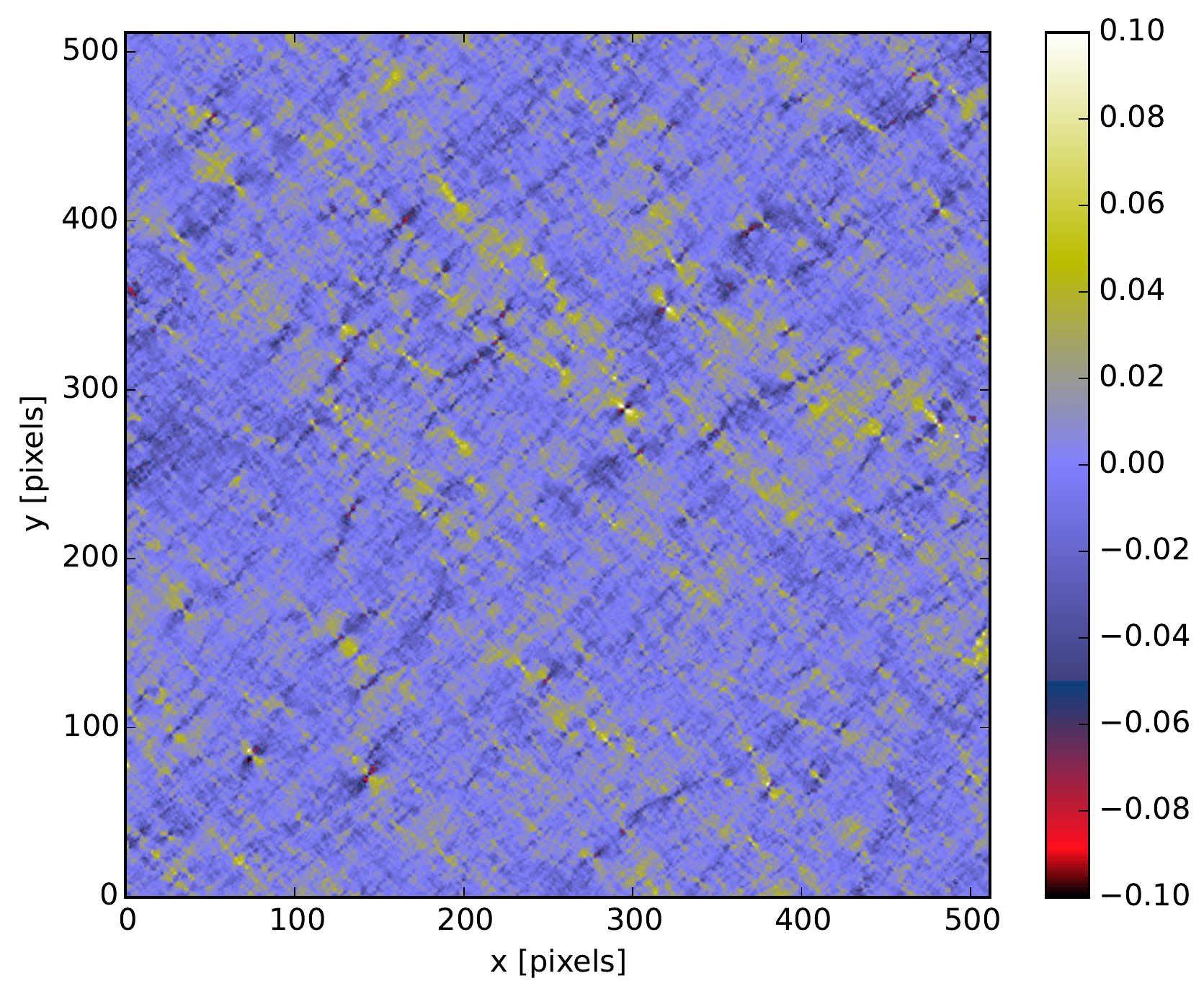,height=6.cm,width=7.cm,clip=}
}}
}
\caption{Simulated shear maps with missing data covering a field of $5^\circ \times 5 ^\circ$. The left panels show the first component of the shear $\gamma_1$ , and the right panels present the second component of the shear $\gamma_2$. The upper panels show the incomplete shear maps, where the pixels with no galaxies are set to zero (displayed in black). The lower panels show the result of the inpainting method that allows us to fill the gaps judiciously.}
\label{shear_mask}
\end{figure*}

In the standard version of the KS method, the pixels with no galaxies are set to zero.
Fig.~\ref{shear_mask} shows an example of simulated shear maps without shape noise derived from the Euclid Flagship mock galaxy catalogue (see Sect. \ref{sect_simu} for more details). The upper panels of Fig.~\ref{shear_mask} show the two components of the shear with zero values (displayed in black) corresponding to the mask of the missing data.
These zero values generate an important leakage during the mass inversion.

With KS+, the problem is reformulated by including additional assumptions to regularise the problem. 
The convergence $\bm{\kappa}$ can be analysed using a transformation $\mathbf \Phi, $ which yields a set of coefficients $\bm \alpha = \mathbf \Phi^{\rm T} \bm{\kappa}$ ($\mathbf \Phi$ is an orthogonal matrix operator, and $\mathbf \Phi^{\rm T}$ represents the transpose matrix of $\mathbf \Phi$).
In the case of the Fourier transformation, $\mathbf \Phi^{\rm T}$ would correspond to the discrete Fourier transform (DFT) matrix, and $\bm \alpha$ would be the Fourier coefficients of $\bm{\kappa}$.
The KS+ method uses a prior of sparsity, that is, it assumes that there is a transformation $\mathbf \Phi$ where the convergence $\bm{\kappa}$ can be decomposed into a set of coefficients $\bm \alpha$, where most of its coefficients are close to zero. 
In this paper, $\mathbf \Phi$ was chosen to be the discrete cosine transform (DCT) following \cite{wl:pires09}. The DCT expresses a signal in terms of a sum of cosine functions with different frequencies and amplitudes. It is similar to the DFT, but uses smoother boundary conditions. This provides a sparser representation. Hence the use of the DCT for JPEG compression.

We can rewrite the relation between the observed shear $\bm{ \gamma}^{\rm{obs} }$ and the noisy convergence $\bm{\kappa}^{\rm n}$ as
\begin{eqnarray}
\bm{\gamma}^{\rm{obs}} =\mathbf M \mathbf{P}  \bm{\kappa}^{\rm n},
\label{miss}
\end{eqnarray}
with $\mathbf M$ being the mask operator and $\mathbf P$ the KS mass-inversion operator.
There is an infinite number of convergence $\bm{\kappa}^{\rm n}$ that can fit the observed shear $\bm \gamma^{\rm{obs}}$.
With KS+, we first impose that the mean convergence vanishes across the survey, as in the KS method. 
Then, among all possible solutions, KS+ searches for the sparsest solution $\tilde{\bm{\kappa}}^{\rm n}$ in the DCT $\mathbf \Phi$ (i.e. the convergence $\bm{\kappa}^{\rm n}$ that can be represented with the fewest large coefficients). 
The solution of this mass-inversion problem is obtained by solving
\begin{equation}
\min_{\bm{\tilde{\kappa}}^{\rm n}}  \| \mathbf \Phi^{\rm T} \bm{\tilde{\kappa}}^{\rm n}  \|_0    \textrm{ subject to }  \parallel  \bm{\gamma}^{\rm{obs}} - \mathbf M  \mathbf P   \bm{\tilde{\kappa}}^{\rm n}   \parallel^2 \le \sigma^2,
\label{eq1}
\end{equation}
where  $|| z ||_0$ the pseudo-norm, that is, the number of non-zero entries in $z$,  $|| z ||$ the classical $l_2$ norm (i.e. $|| z || =\smash{\sqrt{ \sum_{k}(z_{k})^2}}$), and $\sigma$ stands for the  standard deviation of the input shear map measured outside the mask.
The solution of this optimisation task can be obtained through an iterative thresholding algorithm called morphological component analysis (MCA), which was introduced by \cite{mca:elad05} and was adapted to the weak-lensing problem in \cite{wl:pires09}. 

\cite{wl:pires09} used an additional constraint to force the B modes to zero. This is optimal when the shear maps have no B modes. However, any real observation has some residual B modes as a result of intrinsic alignments, imperfect PSF correction, etc. The B-mode power is then transferred to the E modes, which degrades the E-mode convergence reconstruction. We here instead let the B modes free, and an additional constraint was set on the power spectrum of the convergence map. 
To this end, we used a wavelet transform to decompose the convergence maps into a set of aperture mass maps using the starlet transform algorithm \citep{starck:book98,starck:book02}. 
Then, the constraint consists of renormalising the standard deviation (or equivalently, the variance) of each aperture mass map inside the mask regions to the values measured in the data, outside the masks, and then reconstructing the convergence through the inverse wavelet transform.
The variance per scale corresponding to the power spectrum at these scales allows us to constrain a broadband power spectrum of the convergence $\bm \kappa$ inside the gaps.

Adding the power spectrum constraints yields the final sparse optimisation problem,
\begin{equation}
\min_{\bm{\tilde{\kappa}}^{\rm n}}  \| \mathbf \Phi^{\rm T} \bm{\tilde{\kappa}}^{\rm n}  \|_0    \textrm{ s.t. }  \parallel \bm{\gamma}^{\rm{obs}} - \mathbf M \mathbf P \mathbf{W^{\rm T}} \mathbf Q  \mathbf{W} \bm{\tilde{\kappa}}^{\rm n} \parallel^2 \le \sigma^2,
\label{eq2}
\end{equation}
where $\mathbf{W}$ is the forward wavelet transform and $\mathbf{W^{\rm T}}$ its inverse transform, and $\mathbf Q$ is the linear operator used to impose the power spectrum constraint.
More details about the KS+ algorithm are given in Appendix A.

The KS+ method allows us to reconstruct the in-painted convergence maps and the corresponding in-painted shear maps, where the empty pixels are replaced by non-zero values. These interpolated values preserve the continuity of the signal and reduce the signal leakage during the mass inversion (see lower panels of Fig.~\ref{shear_mask}).  
The quality of the convergence maps reconstruction with respect to missing data is evaluated in Sect. \ref{results_1}.
Additionally, the new constraint allows us to use the residual B modes of the reconstructed maps to test for the presence of residual systematic effects and possibly validate the shear measurement processing chain.

\subsection{Field border effects}
\label{border}

\begin{figure*}
\vbox{
\centerline{
\hbox{
\psfig{figure=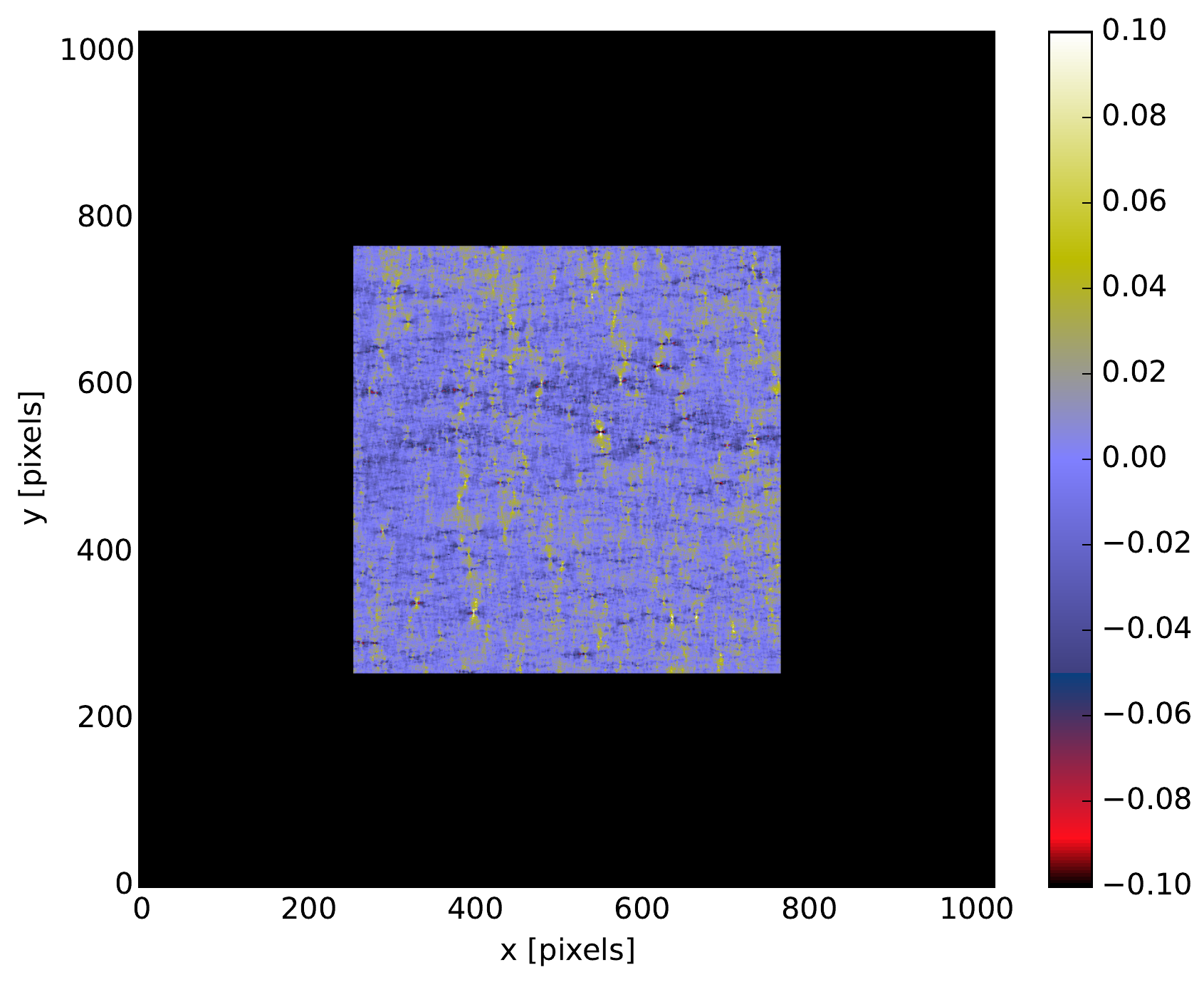,height=6.cm,width=7.cm,clip=}
\hspace{0.2cm}
\psfig{figure=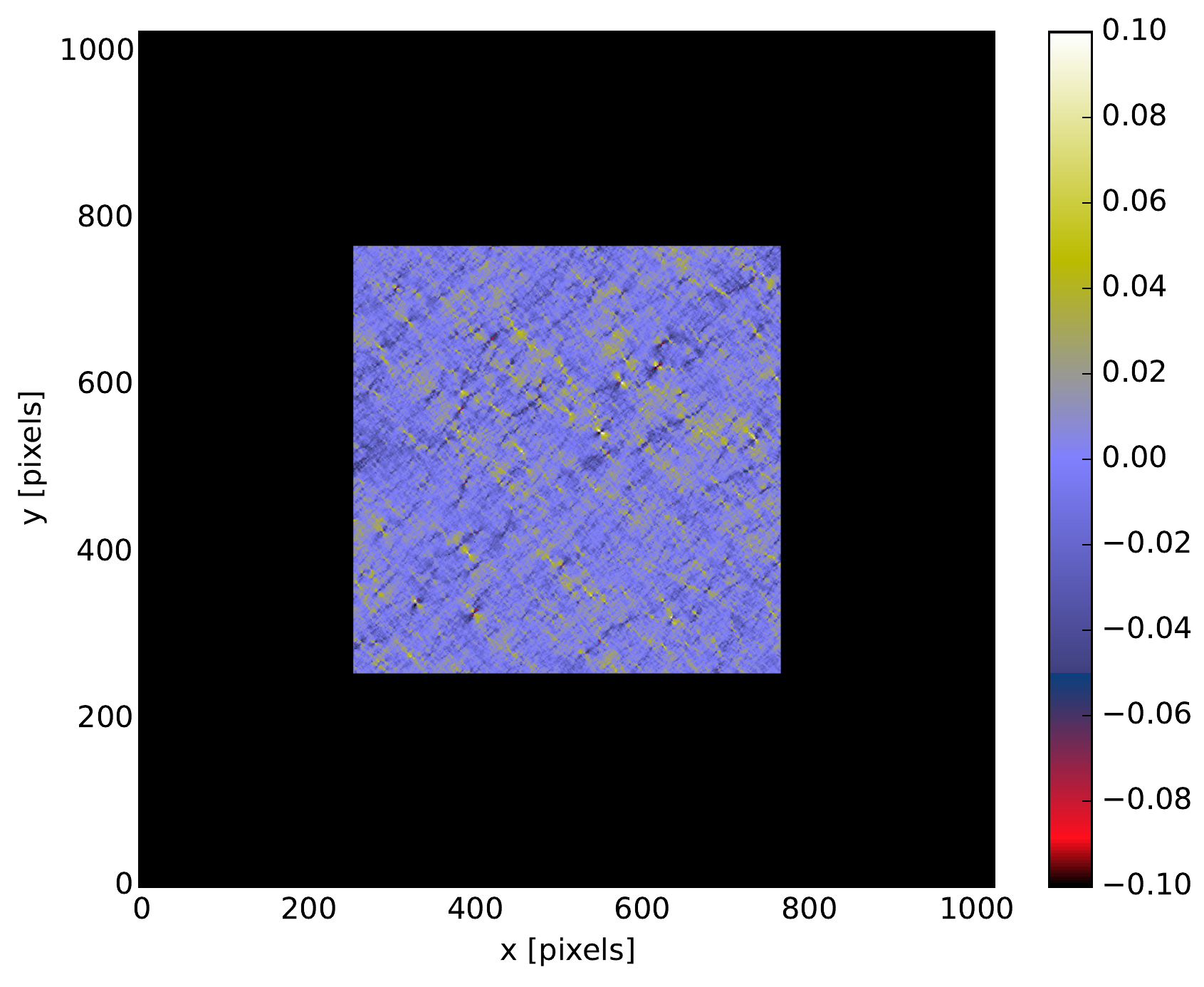,height=6.cm,width=7.cm,clip=}
}}
\vspace{0.3cm}
\centerline{
\hbox{
\psfig{figure=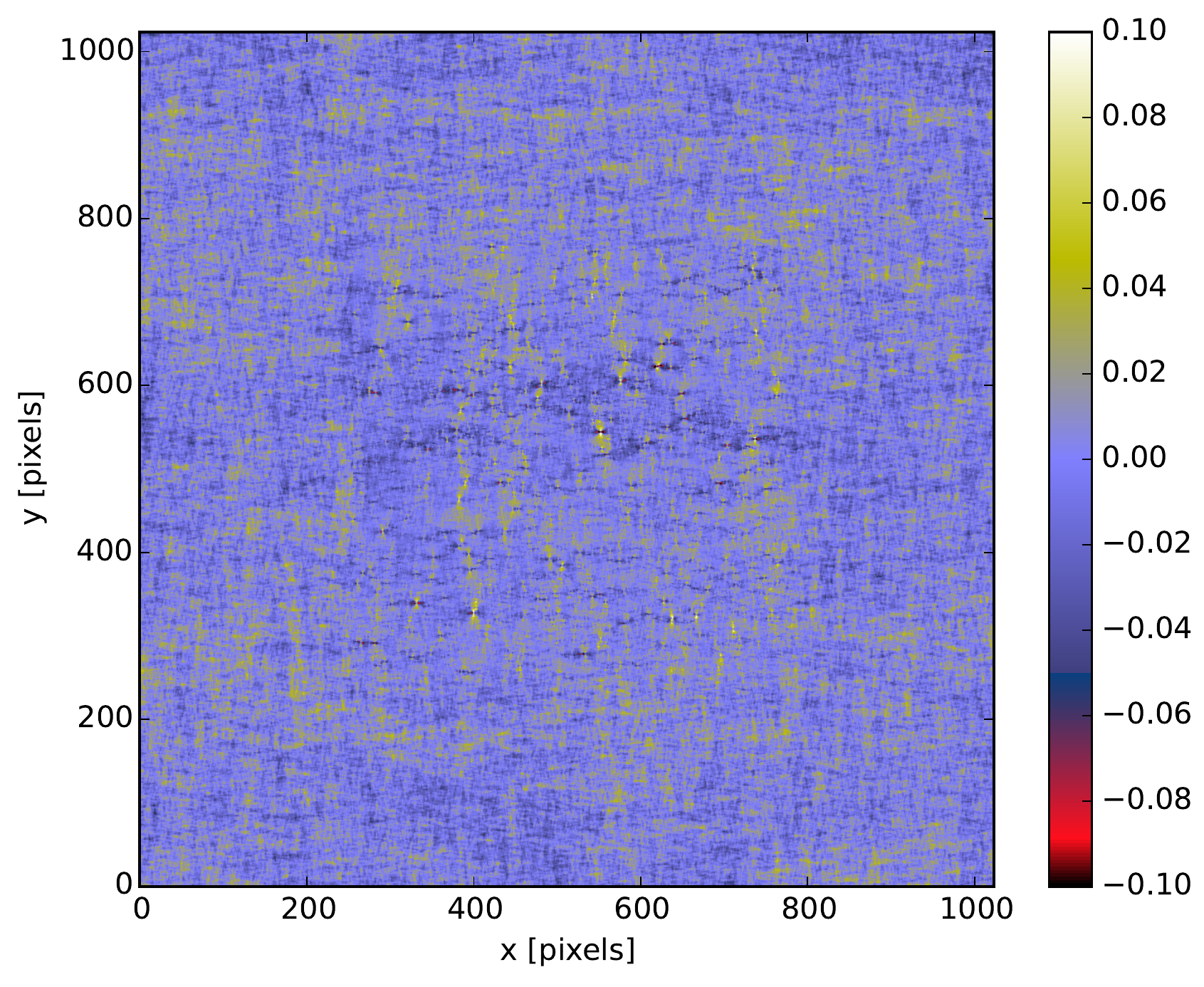,height=6.cm,width=7cm,clip=}
\hspace{0.2cm}
\psfig{figure=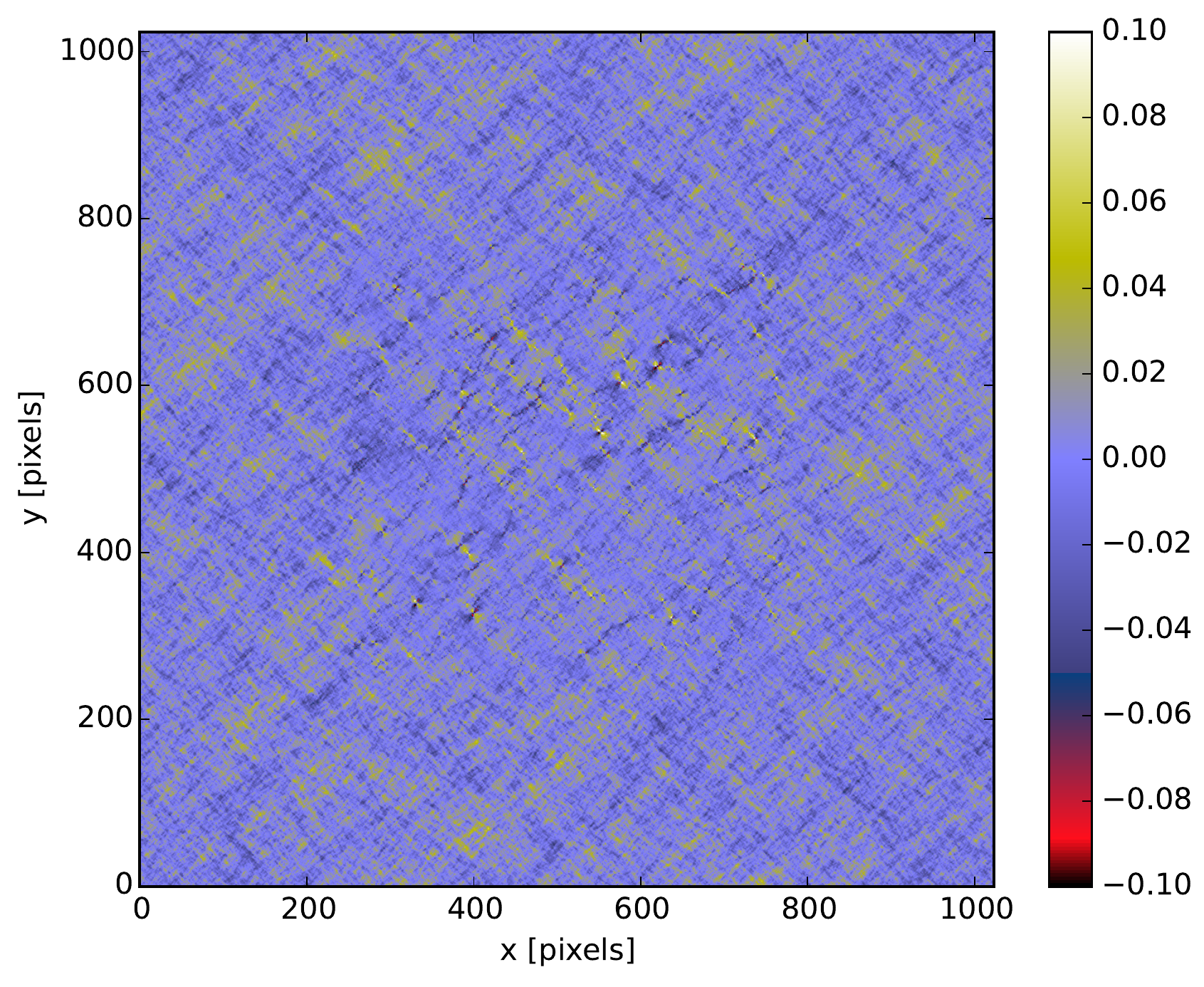,height=6.cm,width=7cm,clip=}
}}
}
\caption{Upper panels: Simulated shear maps covering a field of $5^\circ \times 5 ^\circ$ ,  extended to a field of $10^\circ \times 10^\circ$ by zero padding (zero values are displayed in black). Lower panels: Result of the inpainting method that allows us to extrapolate the shear on the borders. The left panels show the first component of the shear $\gamma_1$, and the right panels present the second component of the shear $\gamma_2$.}
\label{shear_border}
\end{figure*}

The KS and KS+ mass-inversion methods relate the convergence and the shear fields in Fourier space.
However, the discrete Fourier transform implicitly assumes that the image is periodic along both dimensions. Because there is no reason for opposite borders to be alike, the periodic image generally presents strong discontinuities across the frame border. These discontinuities cause several artefacts at the borders of the reconstructed convergence maps. The field border effects can be addressed by removing the image borders, which throws away a large fraction of the data.
Direct finite-field mass-inversion methods have also been proposed \citep[e.g.][]{wl:seitz96,wl:seitz01}. Although unbiased, convergence maps reconstructed using these methods are noisier than those obtained with the KS method.
In the KS+ method, the problem of borders is solved by taking larger support for the image and by considering the borders as masked regions to be in-painted.
The upper panels of Fig.~\ref{shear_border} show the two components of a shear map covering $5^{\circ} \times 5^{\circ}$ degrees and extending to a field of $10^{\circ} \times 10^{\circ}$.
The inpainting method is then used to recover the shear at the field boundaries, as shown in the lower panels of Fig.~\ref{shear_border}. 
After the mass inversion is performed, the additional borders are removed. This technique reduces the field border effects by pushing the border discontinuities farther away.

\subsection{Reduced shear}
\label{reduced}
In Sect. \ref{ks} we assumed knowledge of the shear, in which case the mass inversion is linear.
In practice, the observed galaxy ellipticity is not induced by the shear $\gamma,$ but by the reduced shear $g$ that depends on the convergence $\kappa$ corresponding to that particular line of sight,
\begin{eqnarray}
g \equiv \frac{\gamma}{1-\kappa}.
\label{reducedshear}
\end{eqnarray}
While the difference between the shear $\gamma$ and the reduced shear $g$ is small in the regime of cosmic shear ($\kappa \ll 1$), neglecting it might nevertheless cause a measurable bias at small angular scales \citep[see e.g.][]{wl:white05, wl:shapiro09}.
In the standard version of KS, the Fourier estimators are only valid when the convergence is small ($\kappa \ll 1$),
and they no longer hold near the centre of massive galaxy clusters.
The mass-inversion problem becomes non-linear, and it is therefore important to properly account for reduced shear.

In the KS+ method, an iterative scheme is used to recover the E-mode convergence map, as proposed in \cite{wl:seitz95}. The method consists of solving the linear inverse problem iteratively (see Eq.~\ref{eq:fourier}), using at each iteration the previous estimate of the E-mode convergence to correct the reduced shear using Eq.~(\ref{reducedshear}). 
Each iteration then provides a better estimate of the shear. This iterative algorithm was found in \cite{wl:jullo14} to quickly converge to the solution (about three iterations). The KS+ method uses the same iterative scheme to correct for reduced shear, and we find that it is a reasonable assumption in the case of large-scale structure lensing.

\subsection{Shape noise}

\label{section:shearnoise}

\begin{figure*}
\vbox{
\centerline{
\hbox{
\psfig{figure=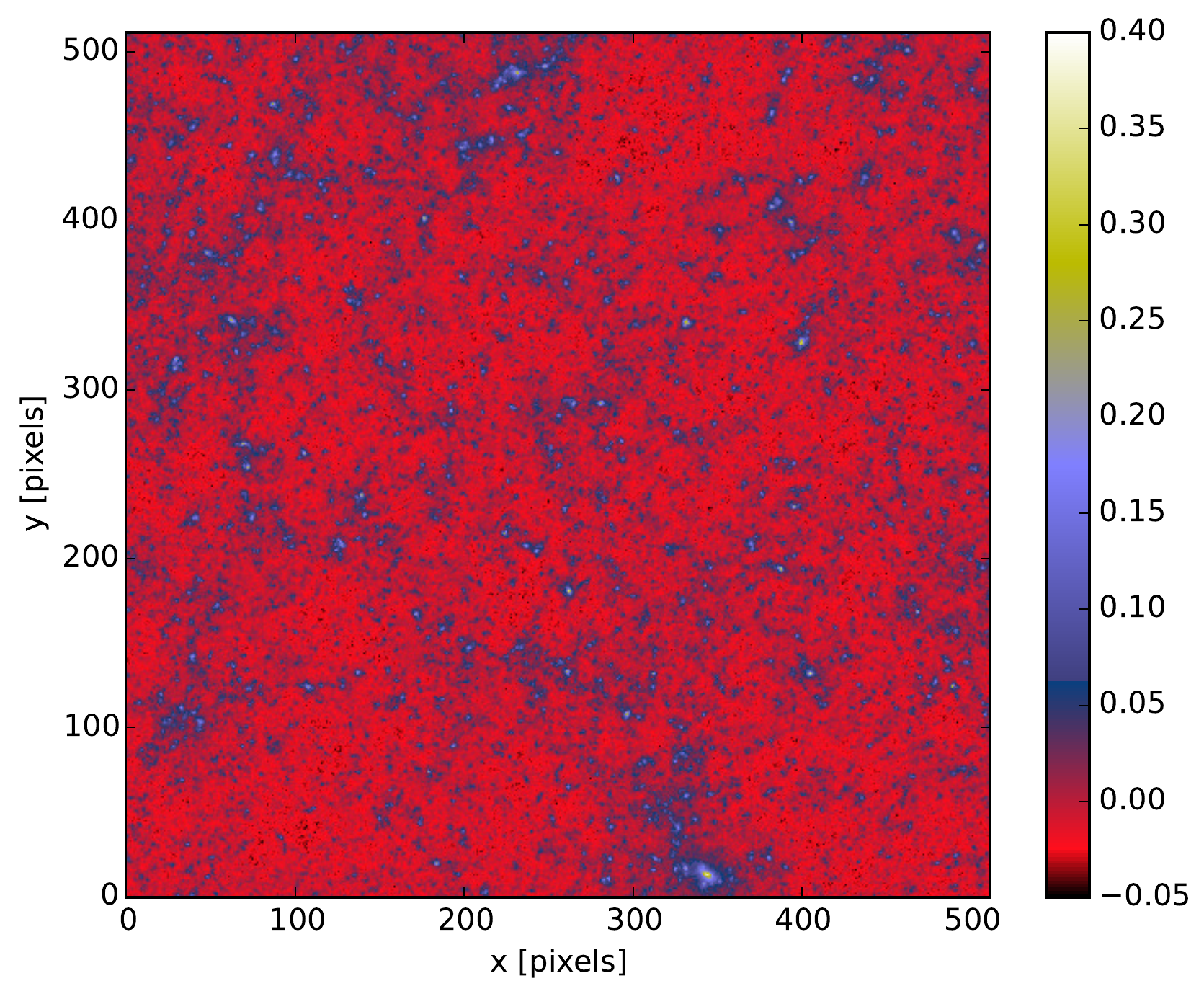,height=6.cm,width=7cm,clip=}
\hspace{0.2cm}
\psfig{figure=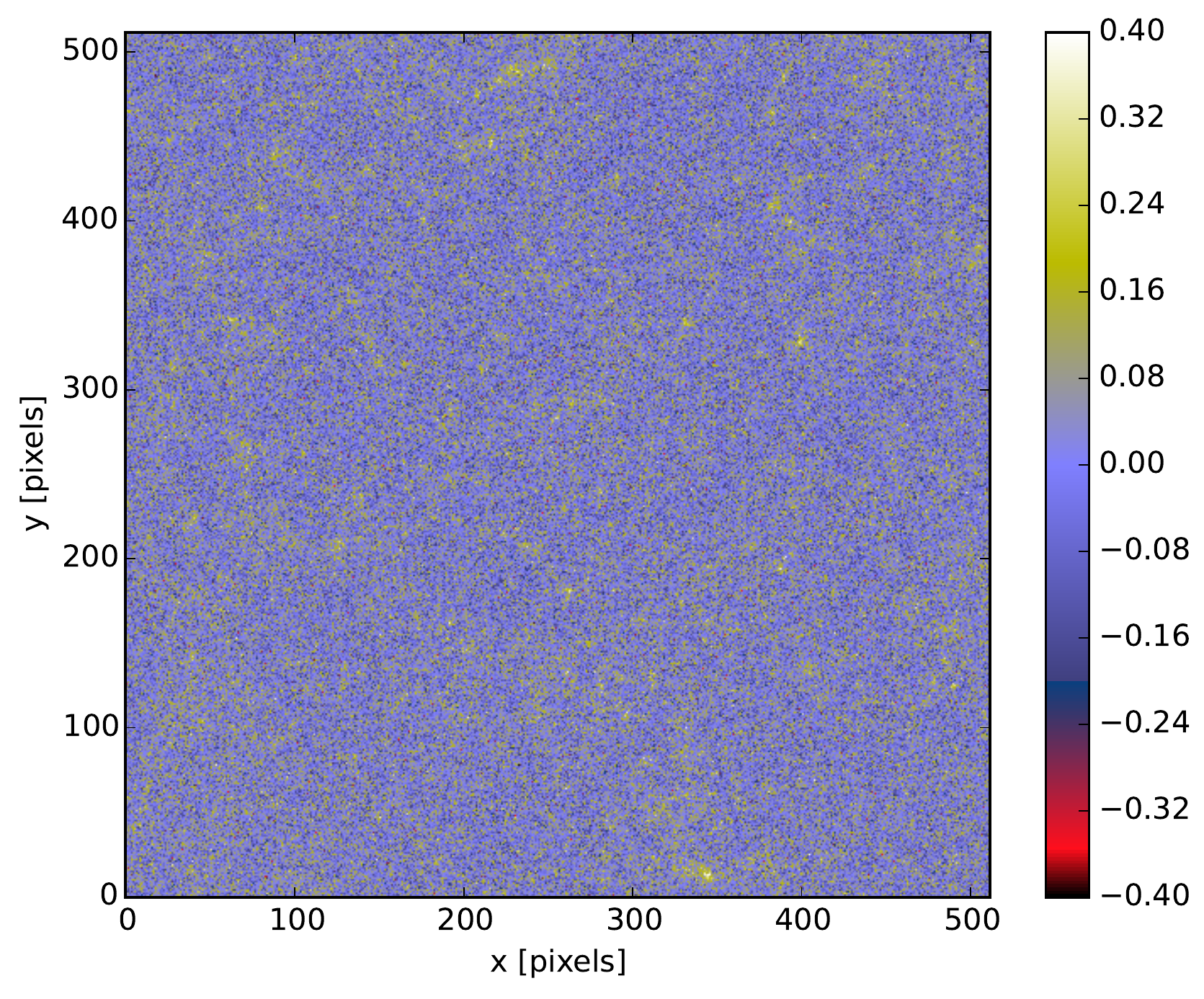,height=6.cm,width=7.cm,clip=}
}}
\vspace{0.3cm}
\centerline{
\hbox{
\psfig{figure=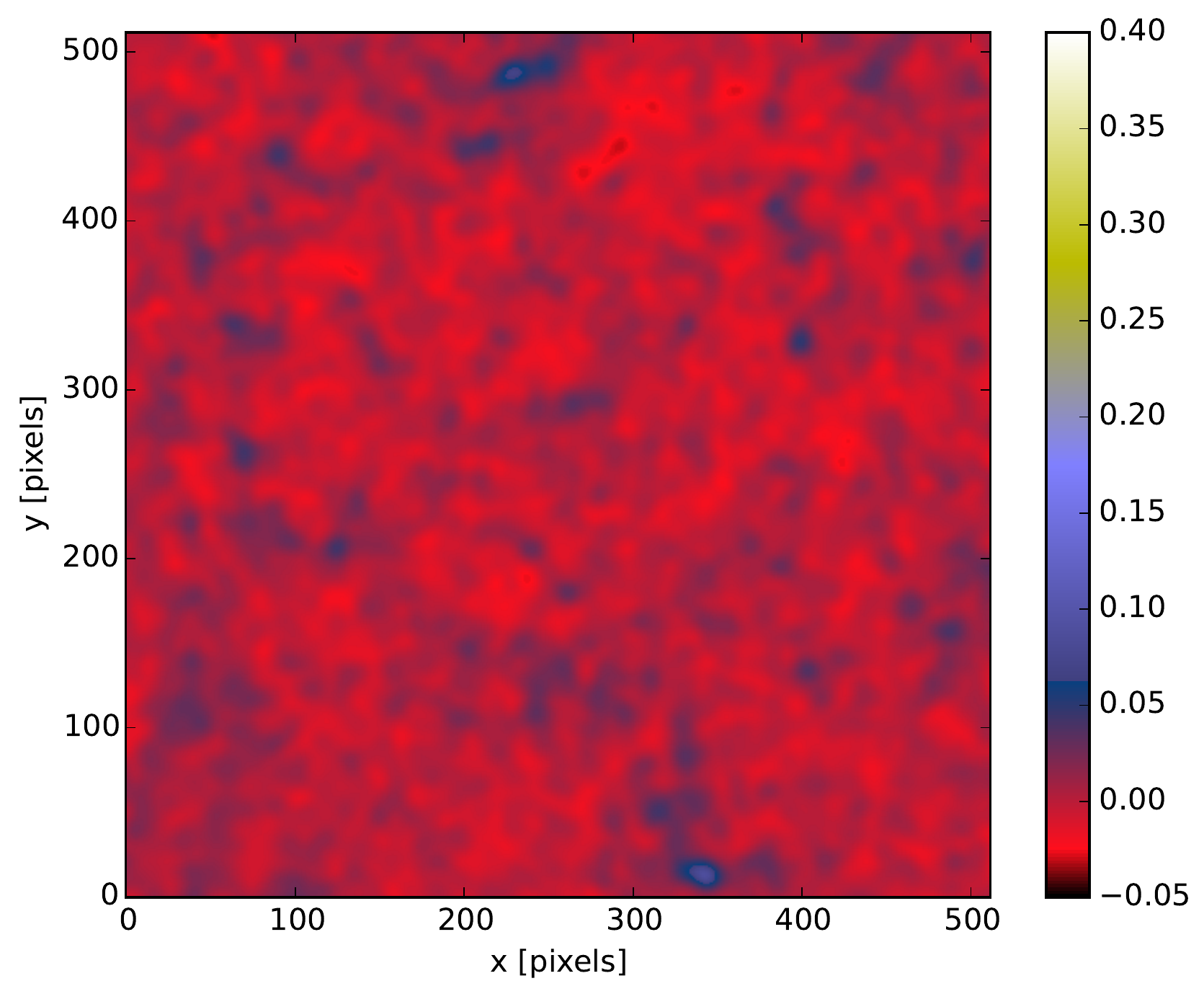,height=6.cm,width=7cm,clip=}
\hspace{0.2cm}
\psfig{figure=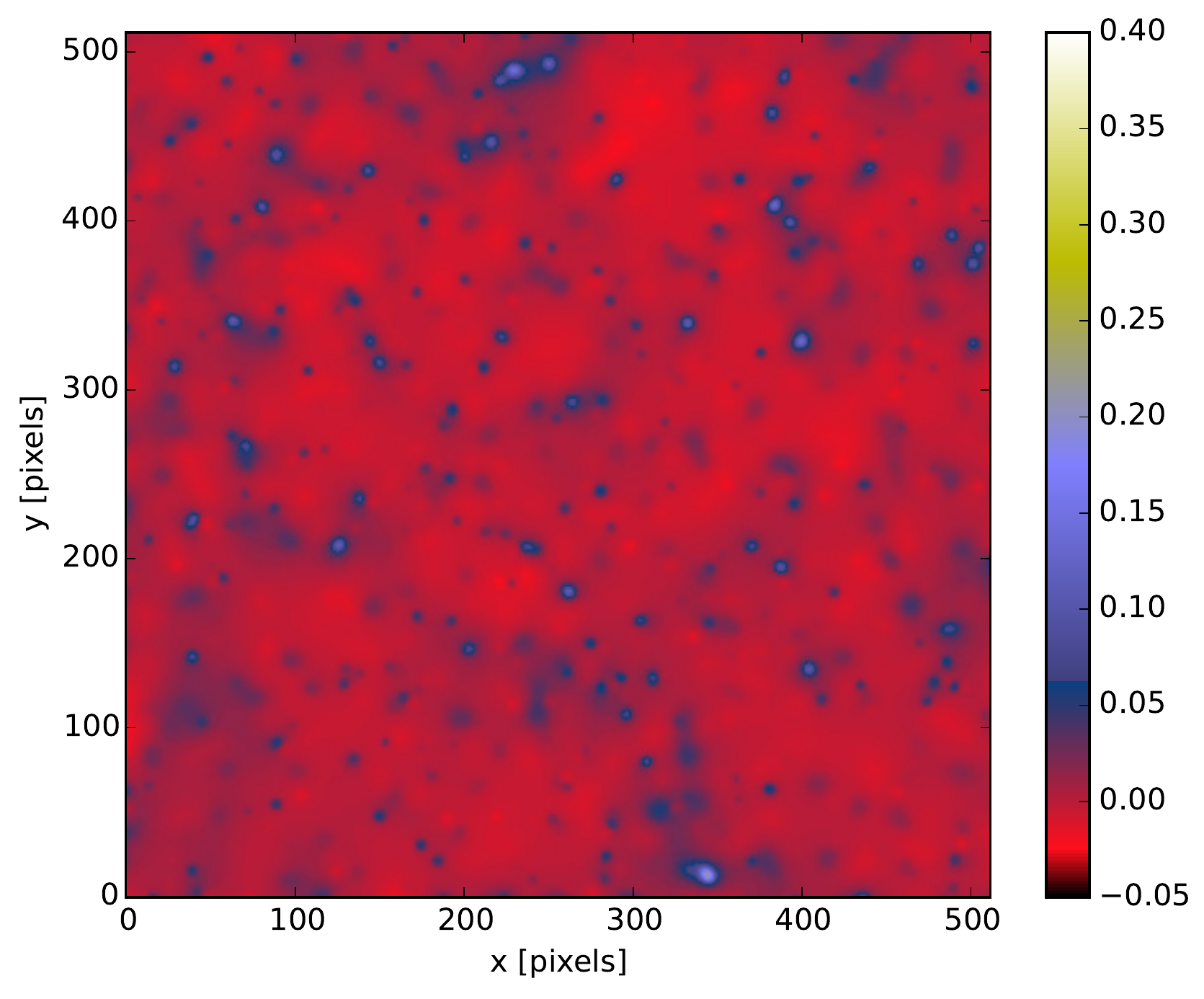,height=6.cm,width=7cm,clip=}
}}

}
\caption{Shape-noise effect. The upper panels show the original E-mode convergence $\kappa$ map (left) and the noisy convergence map with $n_{\rm g} = 30$ gal. arcmin$^{-2}$ (right).  The lower panels show the reconstructed maps using a linear Gaussian filter with a kernel size of $\sigma = 3\arcmin$ (left) and the non-linear MRLens filtering using $\alpha_{\rm FDR} = 0.05$ (right). The field is $5^\circ \times 5^\circ$ downsampled to $512 \times 512$ pixels.}

\label{kappa_noise}
\end{figure*}

In the original implementation of KS, the shear maps are first regularised with a smoothing window (i.e. a low-pass filter) to obtain a smoothed version of the shear field. Then, Eq.~(\ref{eq:fourier}) is applied to derive the convergence maps. 
In contrast, the KS+ method aims at producing very general convergence maps for many applications. In particular, it produces noisy maps with minimum information loss. 

However, for specific applications (e.g. galaxy cluster detection and characterisation), it can be useful to add an additional de-noising step, using any of the many regularisation techniques that have been proposed \citep{wl:bridle98,wl:seitz98,wl:marshall02, wl:starck06, wl:lanusse16}. 
To compare the results of the KS and KS+ methods on noisy maps, we used a linear Gaussian
and the non-linear MRLens filter \citep{wl:starck06} for noise suppression.
Fig.~\ref{kappa_noise} illustrates the effect of shape noise on reconstructing the convergence map.
The upper panels show one E-mode convergence map reconstructed from noise-free (left) and noisy (right) shear data. The convergence map is dominated by the noise. The lower panels show the results of the Gaussian filter (left) and MRLens filter (right). The Gaussian filter gives a smoothed version of the noisy convergence map, whose level of smoothness is set by the width of the Gaussian ($\sigma$). Thus, the amplitude of the over-densities (in blue) are systematically lowered by the Gaussian filter. 
In contrast, 
the MRLens filter uses a prior of sparsity to better recover the amplitude of the structures and uses a parameter, the false-discovery rate ($\alpha_{\rm FDR}$), to control the average fraction of false detections (i.e. the number of pixels that is truly inactive, declared positive) made over the total number of detections \citep{benjamini95}. For some other applications (e.g. two- or three-point correlation), the integrity of the reconstructed noisy convergence maps might be essential and this denoising step can be avoided.

\section{Method}
\label{method}

\subsection{Comparing second-order statistics}
\label{2pcf}
The most common tools for constraining cosmological parameters in weak-lensing studies are the shear two-point correlation functions.
Following \cite{wl:bartelmann01}, they are defined by considering pairs of positions $\vec{\vartheta}$ and $\vec{\theta+\vartheta}$, and defining the tangential and cross-component of the shear $\gamma_{\rm t}$ and $\gamma_{\times}$ at position $\vec{\vartheta}$ for this pair as
\begin{eqnarray}
\gamma_{\rm t} &=& -\operatorname{\mathcal{R}e}(\gamma \operatorname{e}^{-2{\rm i}\varphi}),\\
\gamma_{\times} &=& -\operatorname{\mathcal{I}m}(\gamma \operatorname{e}^{-2{\rm i}\varphi}),
\end{eqnarray}
where $\varphi$ is the polar angle of the separation vector $\vec{\theta}$.
Then we define the two independent shear correlation functions
\begin{eqnarray}
\xi_\pm(\theta) &:=& \langle \gamma_{\rm t} \gamma_{\rm t}' \rangle \pm  \langle \gamma_\times \gamma_\times' \rangle \\
&=& \frac{1}{2\pi} \int_0^{\infty} d\ell \, \ell \, P_{\rm \kappa}(\ell) \,{\rm J}_{0,4}(\ell \theta) ,
\end{eqnarray}
where the Bessel function ${\rm J}_0$ $({\rm J}_4)$ corresponds to the plus (minus) correlation function, $P_{\kappa}(\ell)$ is the power spectrum of the projected matter density, and $\ell$ is the Fourier variable on the sky.
We can also compute the two-point correlation functions of the convergence ($\kappa = \kappa_{\rm E} + \rm i \kappa_{\rm B}$), defined as
\begin{eqnarray}
\xi_{\kappa_{\rm E}}(\theta) = \langle \kappa_{\rm E} \kappa_{\rm E}' \rangle,\nonumber \\
\xi_{\kappa_{\rm B}}(\theta) = \langle \kappa_{\rm B} \kappa_{\rm B}' \rangle.
\end{eqnarray}
We can verify that these two quantities agree \citep{2pcf:schneider02}:
\begin{eqnarray}
 \xi_+(\theta) = \xi_{\kappa_{\rm E}}(\theta) + \xi_{\kappa_{\rm B}}(\theta).
 \end{eqnarray}
When the B modes in the shear field are consistent with zero, the two-point correlation of the shear ($\xi_+$) is equal to the two-point correlation of the convergence $(\xi_{\kappa_{\rm E}})$. Then the differences between the two are due to the errors introduced by the mass inversion to go from shear to convergence.

We computed these two-point correlation functions using the tree code \texttt{athena}  \citep{athena:kilbinger14}. The shear two-point correlation functions were computed by averaging over pairs of galaxies of the mock galaxy catalogue, whereas the convergence two-point correlation functions were computed by averaging over pairs of pixels in the convergence map. The convergence two-point correlation functions can only be computed for separation vectors $\vec{\theta}$ allowed by the binning of the convergence map.

\subsection{Comparing higher-order statistics}
\label{hos}

Two-point statistics cannot fully characterise the weak-lensing field at small scales where it becomes non-Gaussian \citep[e.g.][]{pt:bernardeau02}. Because the small-scale features carry important cosmological information, we computed the third-order moment, $\langle \kappa_{\rm E}^3 \rangle$, and the fourth-order moment,  $\langle \kappa_{\rm E}^4\rangle$, of the convergence. Computations were performed on the original convergence maps provided by the Flagship simulation, as well as on the convergence maps reconstructed from the shear field with the KS and KS+ methods. 
We evaluated the moments of convergence at various scales by computing aperture mass maps \citep{map:schneider96, map:schneider97}. Aperture mass maps are typically obtained by convolving the convergence maps with a filter function of a specific scale (i.e. aperture radii). We performed this here by means of a wavelet transform using the starlet transform algorithm \citep{starck:book98,starck:book02}, which simultaneously produces a set of aperture mass maps on dyadic (powers of two) scales (see Appendix A for more details). Leonard et al. (2012) demonstrated that the aperture mass is formally identical to a wavelet transform at a specific scale and the aperture mass filter corresponding to this transform is derived. The wavelet transform offers significant advantages over the usual aperture mass algorithm in terms of computation time, providing speed-up factors of about 5 to 1200 depending on the scale.

\subsection{Numerical simulations}
\label{sect_simu}

We used the Euclid Flagship mock galaxy catalogue version 1.3.3
(Castander F. et al., in prep) derived from N-body cosmological simulation \citep{flagship:potter17} with parameters $\Omega_{\rm m}=0.319$, $\Omega_{\rm b} = 0.049$, $\Omega_{\Lambda} = 0.681$, $\sigma_8=0.83$, $n_{\rm s}=0.96$, $h=0.67$, and the particle mass was \smash{$m_{\rm p} \sim 2.398 \times10^9\,\text{\ensuremath{\textup{M}_{\odot}}} h^{-1}$}. The galaxy light-cone catalogue contains 2.6 billion galaxies over $5000\,\deg^2$ , and it extends up to $z=2.3$. It has been built using a hybrid halo occupation distribution and halo abundance matching (HOD+HAM) technique, whose galaxy-clustering properties were discussed in detail in \cite{flagship:crocce15}.
The lensing properties were computed using the Born approximation and projected mass density maps (in \texttt{HEALPix} format with $N_{\rm side}=8192$) generated from the particle light-cone of the Flagship simulation.
More details on the lensing properties of the Flagship mock galaxy catalogue can be found in \cite{flagship:fosalba15,flagship:fosalba18}.

In order to evaluate the errors introduced by the mass-mapping methods, we extracted ten contiguous shear and convergence fields of $10^\circ \times 10^\circ$ from the Flagship mock galaxy catalogue, yielding a total area of 1000 deg$^2$. The fields correspond to galaxies that lie in the range of \smash{$15^\circ < \alpha < 75^\circ$} and \smash{$15^\circ < \delta < 35^\circ$} , where $\alpha$ and $\delta$ are the right ascension and declination, respectively.
In order to obtain the density of 30 galaxies per arcmin$^2$ foreseen for the Euclid Wide survey, we randomly selected one quarter of all galaxies in the catalogue. 
Then projected shear and convergence maps were constructed by combining all the redshifts of the selected galaxies.
More sophisticated selection methods based on galaxy magnitude would produce slightly different maps. However, they would not change the performances of the two methods we studied here.
The fields were down-sampled to $1024 \times 1024$ pixels, which corresponds to a pixel size of about \ang[angle-symbol-over-decimal]{;0.6;}. Throughout all the paper, the shaded regions stand for the uncertainties on the mean estimated from the total 1000 deg$^2$ of the ten fields. Because the Euclid Wide survey is expected to be 15 000 deg$^2$, the sky coverage will be 15 times larger than the current mock. Thus, the uncertainties will be smaller by a factor of about 4.

\begin{figure*}
\vbox{
\centerline{
\hbox{
\psfig{figure=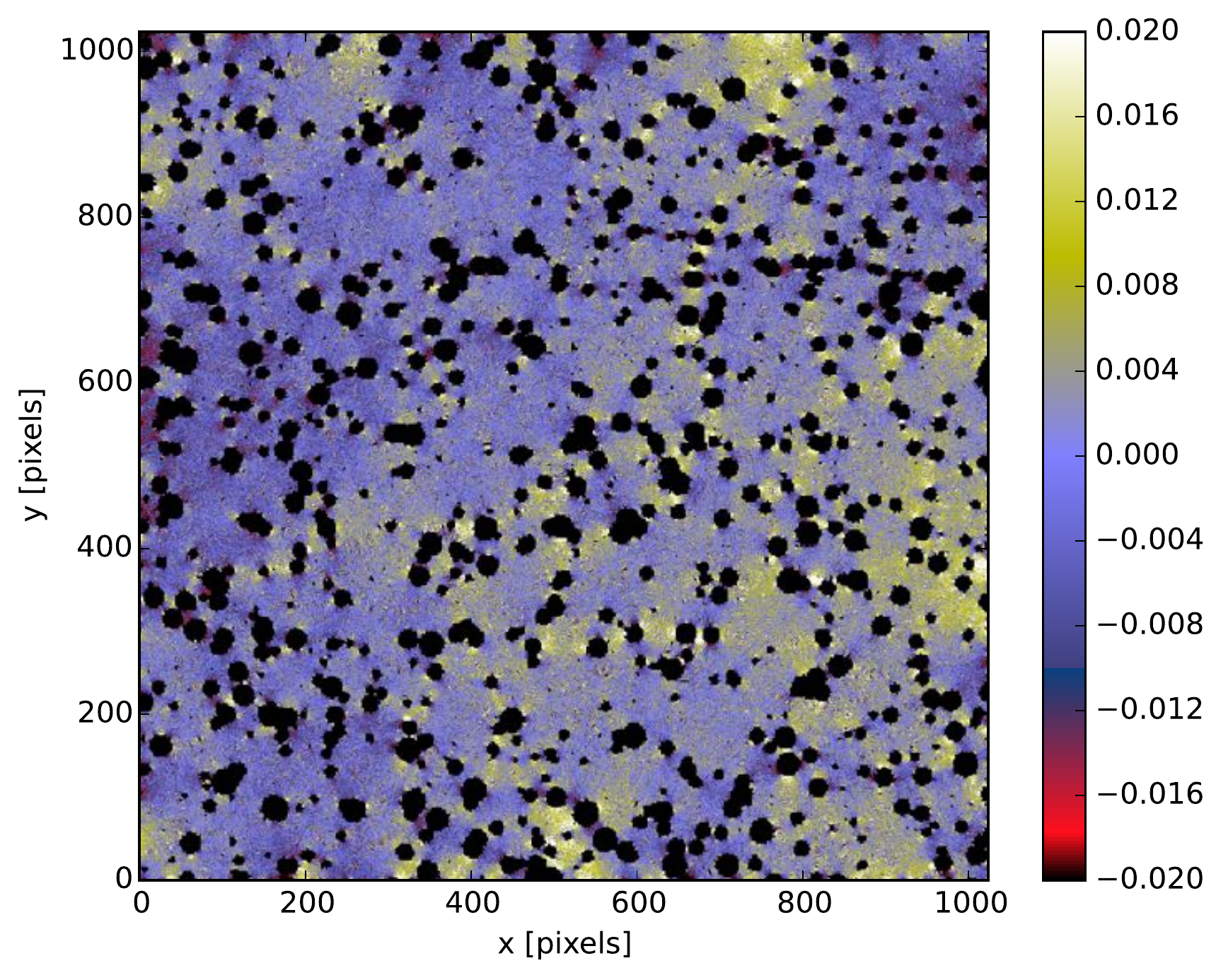,height=6.cm,width=7cm,clip=}
\hspace{0.2cm}
\psfig{figure=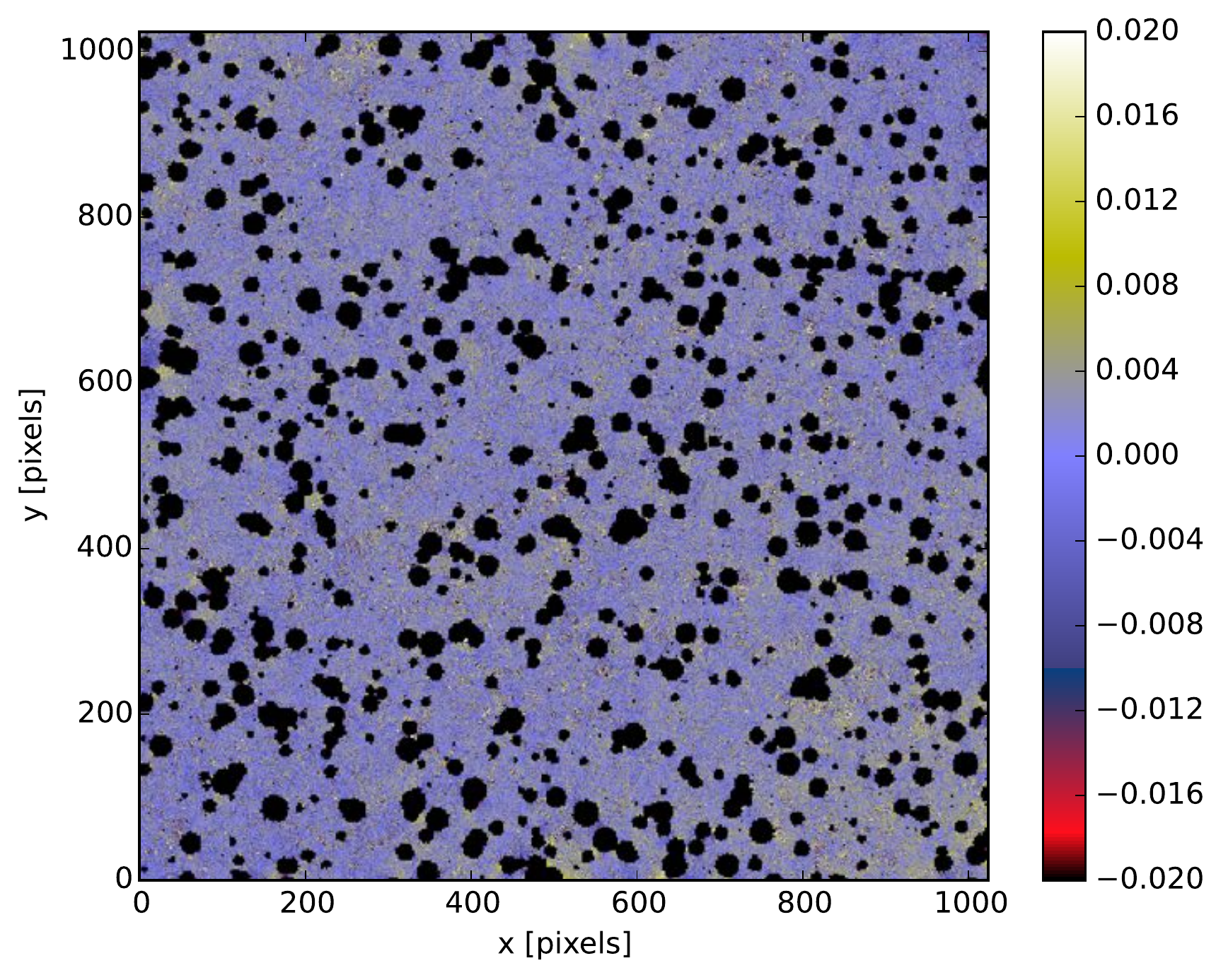,height=6.cm,width=7cm,clip=}
}}
}
\caption{Missing data effects: Pixel difference outside the mask between the original E-mode convergence $\kappa$ map and the map reconstructed from the incomplete simulated noise-free shear maps using the KS method (left) and the KS+ method (right). The field is $10^\circ \times 10^\circ$ downsampled to $1024 \times 1024$ pixels. The missing data represent roughly 20\% of the data.}
\label{kappa_missing}
\end{figure*}

\subsection{Shear field projection}
\label{projections}

We considered fields of $10^\circ \times10^\circ$. The fields were taken to be sufficiently small to be approximated by a tangent plane. 
We used a gnomonic projection to project the points of the celestial sphere onto a tangent plane, following \cite{cmb:pires12}, who found that this preserves the two-point statistics. We note, however, that higher-order statistics may behave differently under different projections.

The shear field projection is obtained by projecting the galaxy positions from the sphere ($\alpha$, $\delta$) in the catalogue onto a tangent plane ($x$, $y$).
The projection of a non-zero spin field such as the shear field requires a projection of both the galaxy positions and their orientations.
Projections of the shear do not preserve the spin orientation, which can generate substantial B modes (depending on the declination) if not corrected for.  
Two problems must be considered because of the orientation. First, the projection of the meridians are not parallel, so that north is not the same everywhere in the same projected field of view. Second, the projection of the meridians and great circles is not perpendicular, so that the system is locally non-Cartesian.
Because we properly correct for the other effects (e.g. shape noise, missing data, or border effects) and consider large fields of view ($10^{\circ} \times 10^{\circ}$) possibly at high latitudes, these effects need to be considered.
The first effect is dominant and generates substantial B modes (increasing with latitude) if not corrected for. This can be easily corrected 
for by measuring the shear orientation with respect to local north. We find that this correction is sufficient for the residual errors due to projection to become negligible compared to errors due to other effects.

\section{Systematic effects on the mass-map inversion}
\label{results_1}
In this section, we quantify the effect of field borders, missing data, shape noise, and the approximation of shear by reduced shear on the KS and KS+ mass-inversion methods. The quality of the reconstruction is assessed by comparing the two-point correlation functions, third- and fourth-order moments.

\subsection{Missing data effects}
\label{sect_missing}

We used the ten noise-free shear fields of $10^\circ \times 10^\circ$ described in Sect.~\ref{sect_simu} and the corresponding noise-free convergence maps.
We converted the shear fields into planar convergence maps using the KS and KS+ methods, masking 20\% of the data as expected for the \Euclid survey. 
The mask was derived from the Data Challenge 2 catalogues produced by the Euclid collaboration using the code \texttt{FLASK} \citep{flask:xavier16}.

\begin{figure}
\centerline{
\psfig{figure=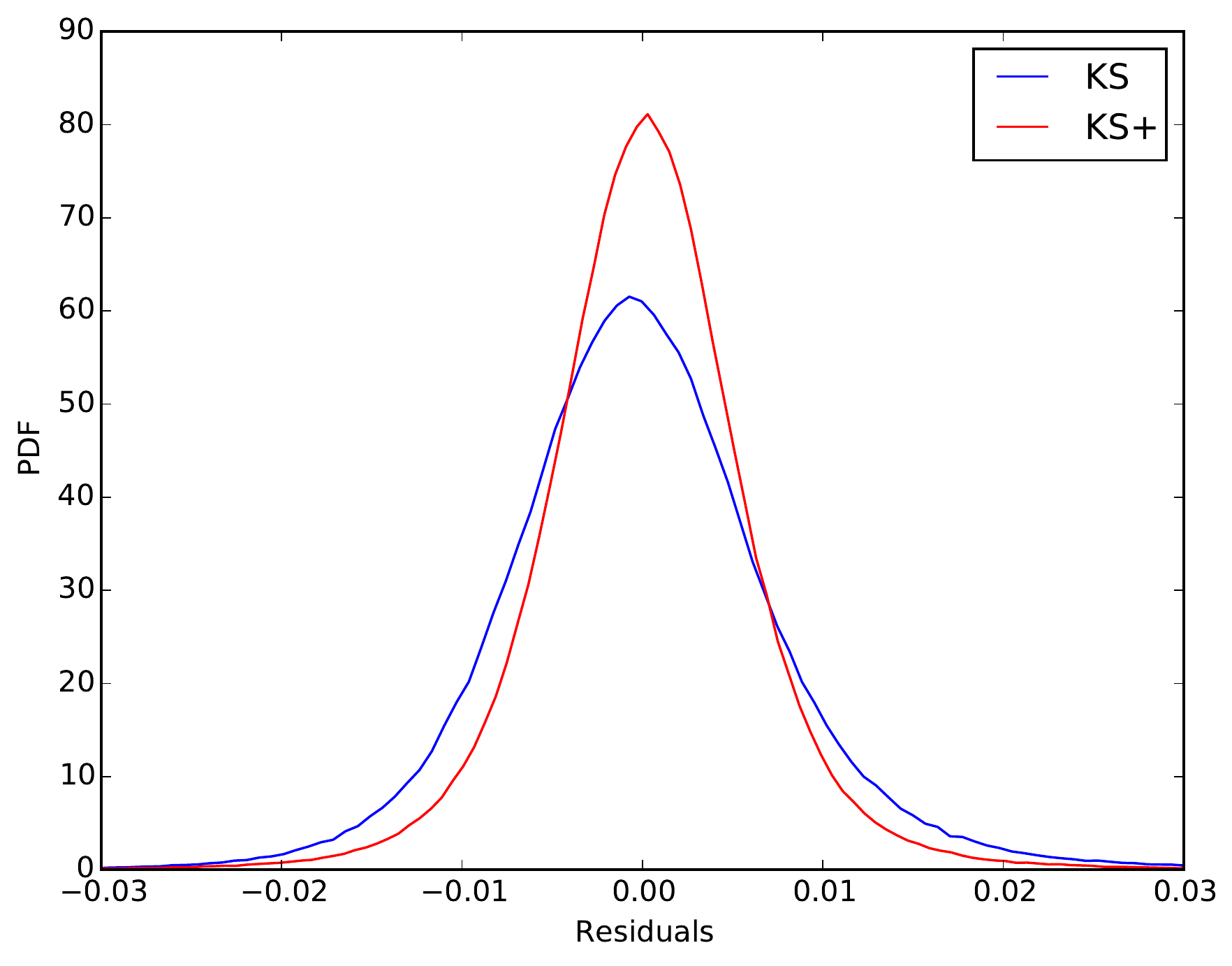,height=6.5cm,clip=}
}
\caption{Missing data effects: PDF of the residual errors between the original E-mode convergence map and the reconstructed maps using KS (blue) and KS+ (red), measured outside the mask.}
\label{missing_residual_PDF}
\end{figure}

Fig.~\ref{kappa_missing} compares the results of the KS and KS+ methods in presence of missing data.
The figure shows the residual maps, that is, the pixel difference between the original E-mode convergence map and the reconstructed maps.
The amplitude of the residuals is larger with the KS method. Detailed investigation shows that the excess error is essentially localised around the gaps. 
Because the mass inversion operator $\mathbf P$ is intrinsically non-local, it generates artefacts around the gaps.
In order to quantify the average errors,
Fig.~\ref{missing_residual_PDF} shows the probability distribution function (PDF) of the residual maps, estimated outside the mask. 
The standard deviation is 0.0080 with KS and 0.0062 with KS+. The residual errors obtained with KS are then 30\% larger than those obtained with KS+.

\begin{figure}[h!]
\vbox{
\centerline{
\psfig{figure=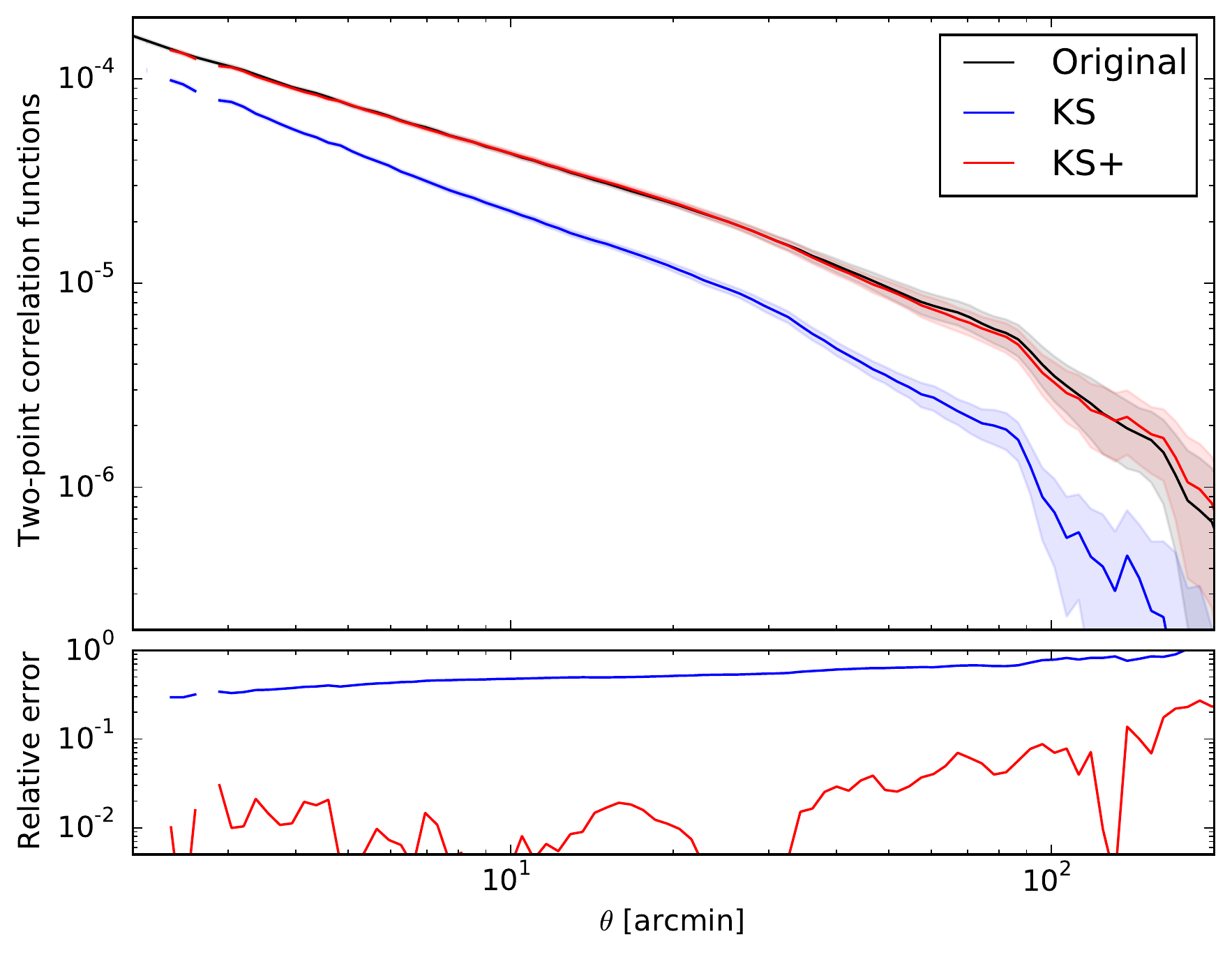,width=9.cm,clip=}
}
\caption{Missing data effects: Mean shear two-point correlation function $\xi_+$ (black) and corresponding mean convergence two-point correlation function $\xi_{\kappa_{\rm E}}$ reconstructed using the KS method (blue) and using the KS+ method (red) from incomplete shear maps. The estimation is only made outside the mask $M$. The shaded area represents the uncertainties on the mean estimated on 1000 deg$^{2}$. The lower panel shows the relative two-point correlation errors introduced by missing data effects, that is, the normalised difference between the upper curves.}
\label{missing_corr}
}
\end{figure}

\begin{figure}
\vbox{
\centerline{
\hbox{
\psfig{figure=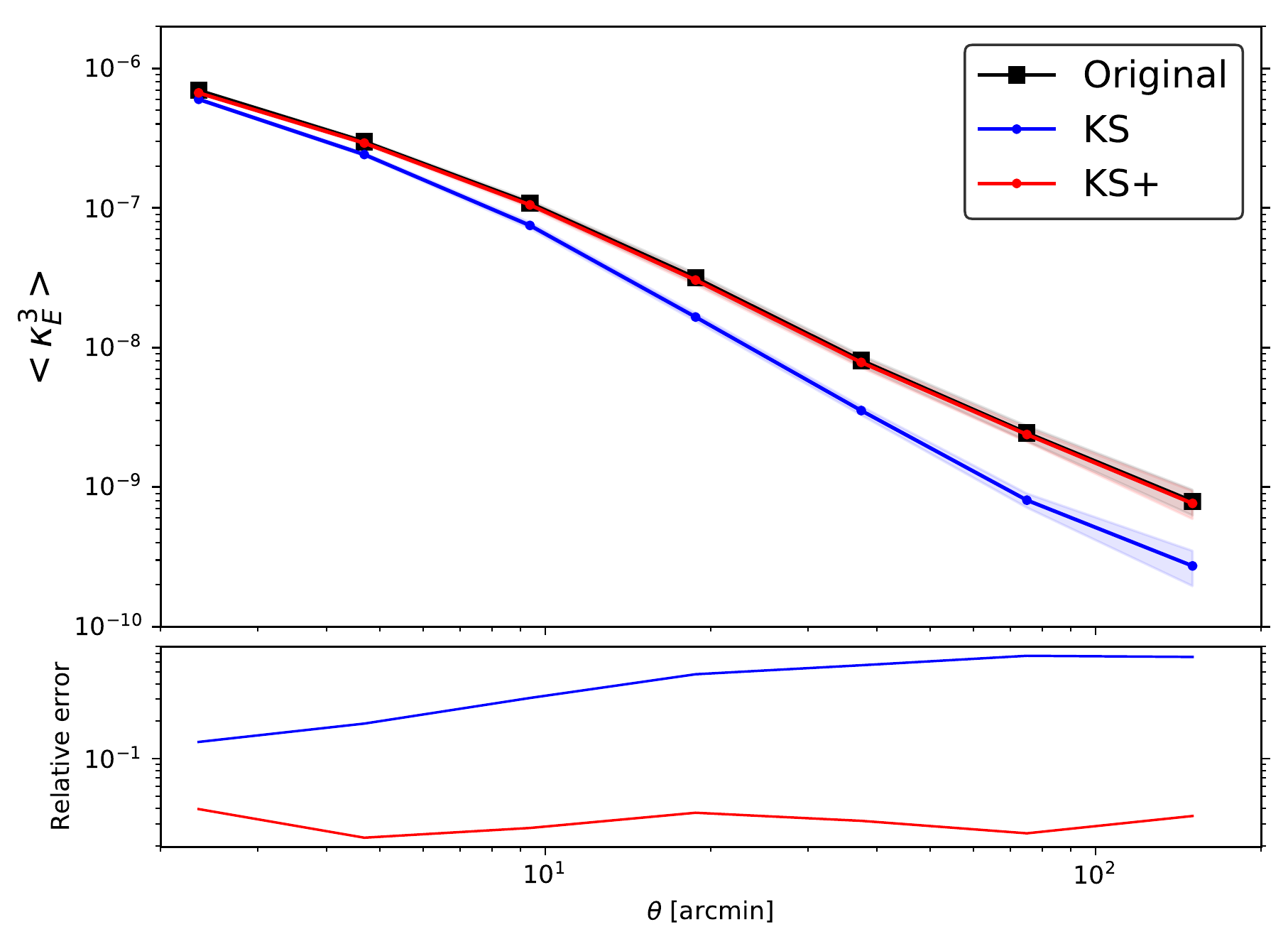,width=9.cm,clip=}
}}
\vspace{0.3cm}
\centerline{
\hbox{
\psfig{figure=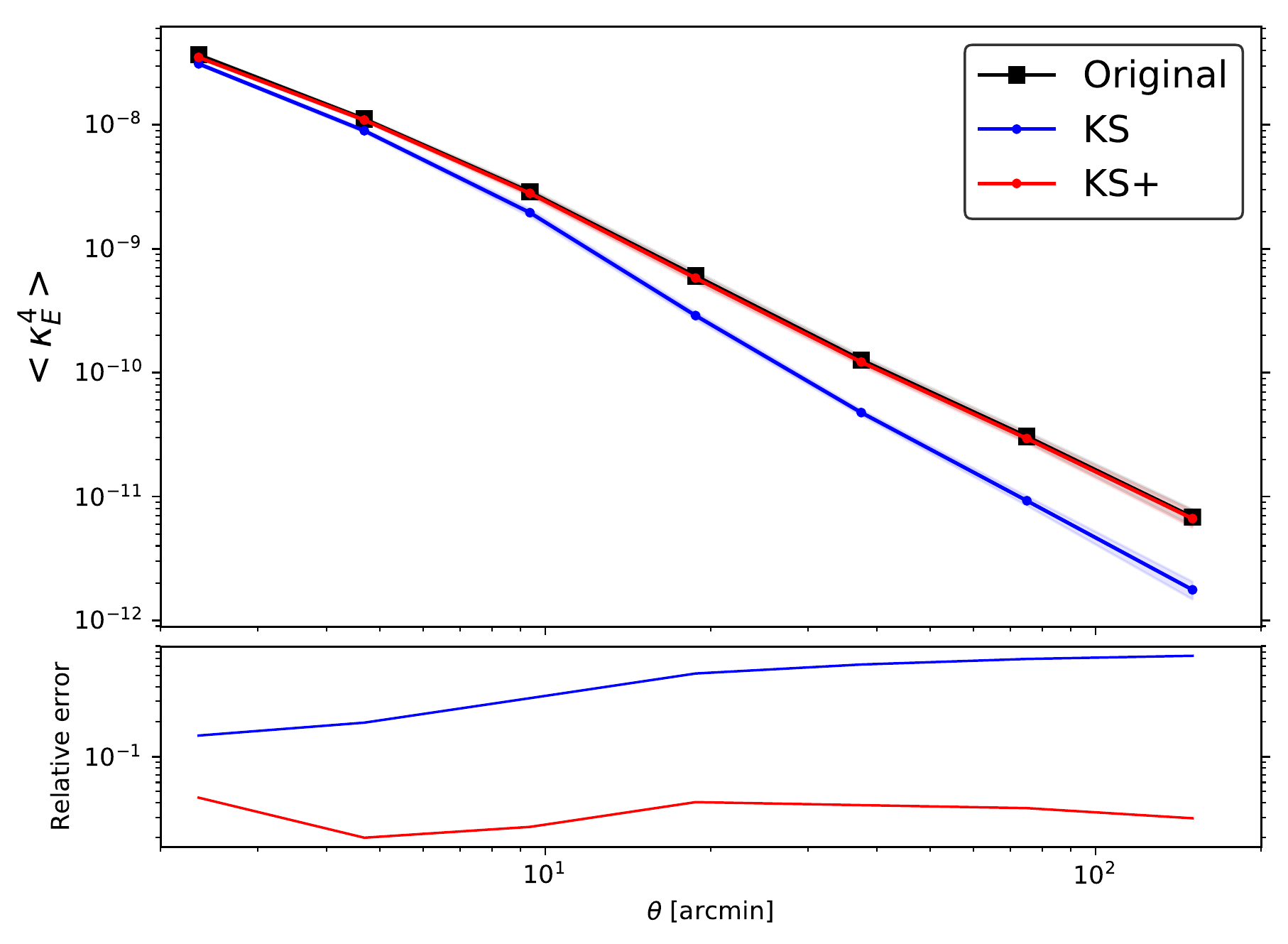,width=9.cm,clip=}
}}
}
\caption{Missing data effects: Third-order (upper panel) and fourth-order (lower panel) moments estimated on seven wavelet bands of the original E-mode convergence map (black) compared to the moments estimated on the KS (blue) and KS+ (red) convergence maps at the same scales. The KS and KS+ convergence maps are reconstructed from incomplete noise-free shear maps. The estimation of the third- and fourth-order moments is made outside the mask. The shaded area represents the uncertainties on the mean estimated on 1000 deg$^{2}$. The lower panel shows the relative higher-order moment errors introduced by missing data effects, that is, the normalised difference between the upper curves.}
\label{missing_hos}
\end{figure}


The quality of the mass inversion at different scales can be estimated using the two-point correlation function and higher-order moments computed at different scales.
Fig.~\ref{missing_corr} compares the two-point correlation functions computed on the convergence and shear maps outside the mask.
Because the B mode is consistent with zero in the simulations, we expect that these two quantities are equal within the precision of the simulations (see Sect.~\ref{2pcf}).
The KS method systematically underestimates the original two-point correlation function by a factor of about 2 on arcminute scales, but can reach factors of 5 at larger scales.
The mass-inversion operator $\mathbf P$ being unitary, the signal energy is conserved by the transformation (i.e. $\sum(\gamma_1^2+ \gamma_2^2) = \sum(\kappa_{\rm E}^2+ \kappa_{\rm B}^2)$, where the summation is performed over all the pixels of the maps). We found that about 10\% of the total energy leaks into the gaps and about 15\% into the B-mode component.
In contrast, the errors of the KS+ method are of the order of a few percent at scales smaller than $1^\circ$. At any scale, the KS+ errors are about 5-10 times smaller than the KS errors, remaining in the $1\sigma$ uncertainty of the original two-point correlation function.

Fig.~\ref{missing_hos} shows the third-order (upper panel) and fourth-order (lower panel) moments estimated at six different wavelet scales ( \ang[angle-symbol-over-decimal]{;2.34;},  \ang[angle-symbol-over-decimal]{;4.68;},  \ang[angle-symbol-over-decimal]{;9.37;},  \ang[angle-symbol-over-decimal]{;18.75;},  \ang[angle-symbol-over-decimal]{;37.5;}, and \ang[angle-symbol-over-decimal]{;75.0;}) using the KS and KS+methods. For this purpose, the pixels inside the mask were set to zero in the reconstructed convergence maps. The aperture mass maps corresponding to each wavelet scale were computed, and the moments were calculated outside the masks.

The KS method systematically underestimates the third- and fourth-order moments at all scales.
Below 10$\arcmin$, the errors on the moments remain smaller than 50\%, and they increase with scale up to a factor 3. In comparison, the KS+ errors remain much smaller at all scales, and remain within the 1$\sigma$ uncertainty.

\subsection{Field border effects}
\begin{figure*}
\centerline{
\hbox{
\psfig{figure=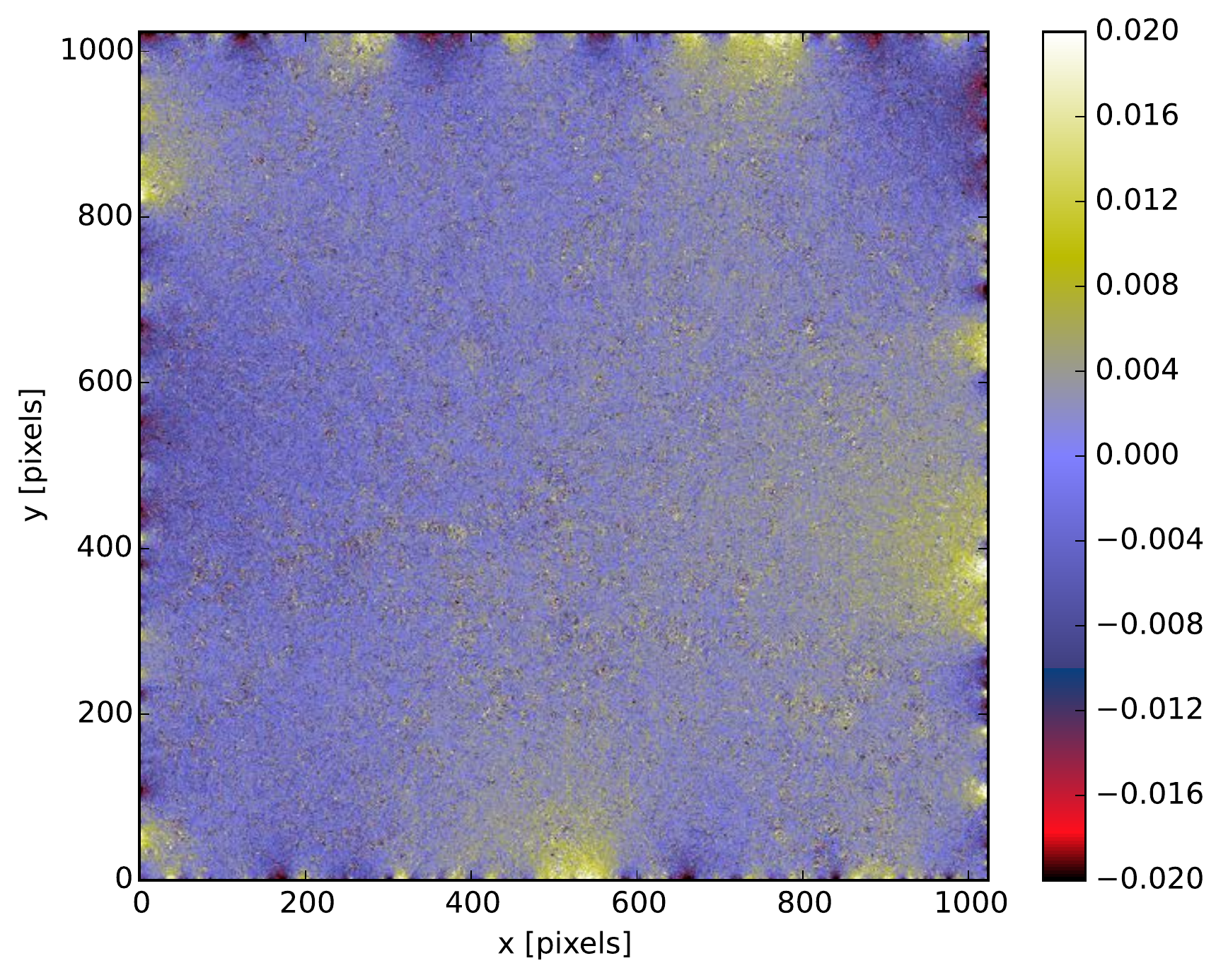,height=6.cm,width=7cm,clip=}
\hspace{0.2cm}
\psfig{figure=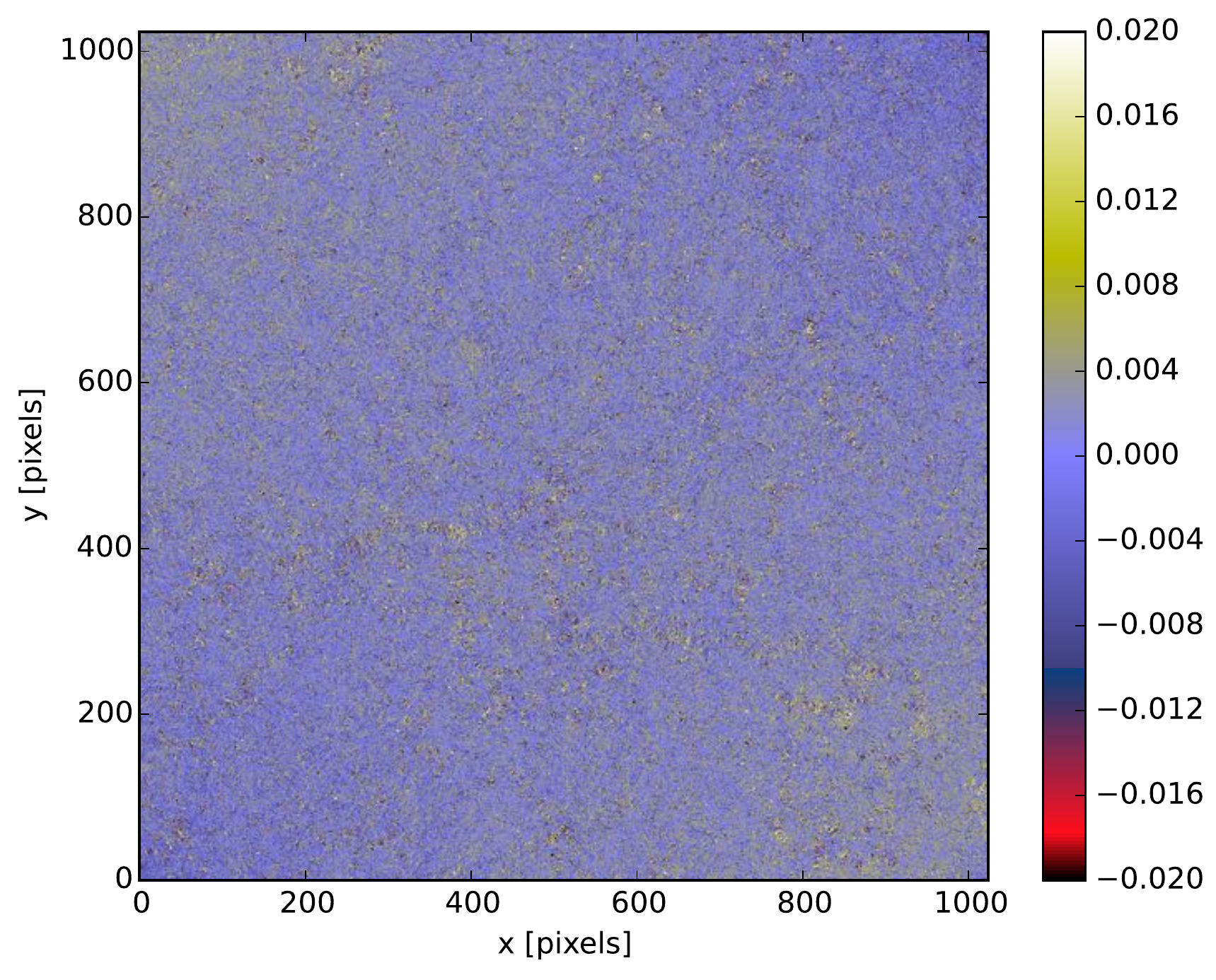,height=6.cm,width=7cm,clip=}
}}
\caption{Field border effects: Pixel difference between the original E-mode convergence $\kappa$ map and the map reconstructed from the corresponding simulated shear maps using the KS method (left) and the KS+ method (right). The field is $10^\circ \times 10^\circ$ downsampled to $1024 \times 1024$ pixels.}
\label{kappa_borders}
\end{figure*}

\begin{figure}
\centerline{
\psfig{figure=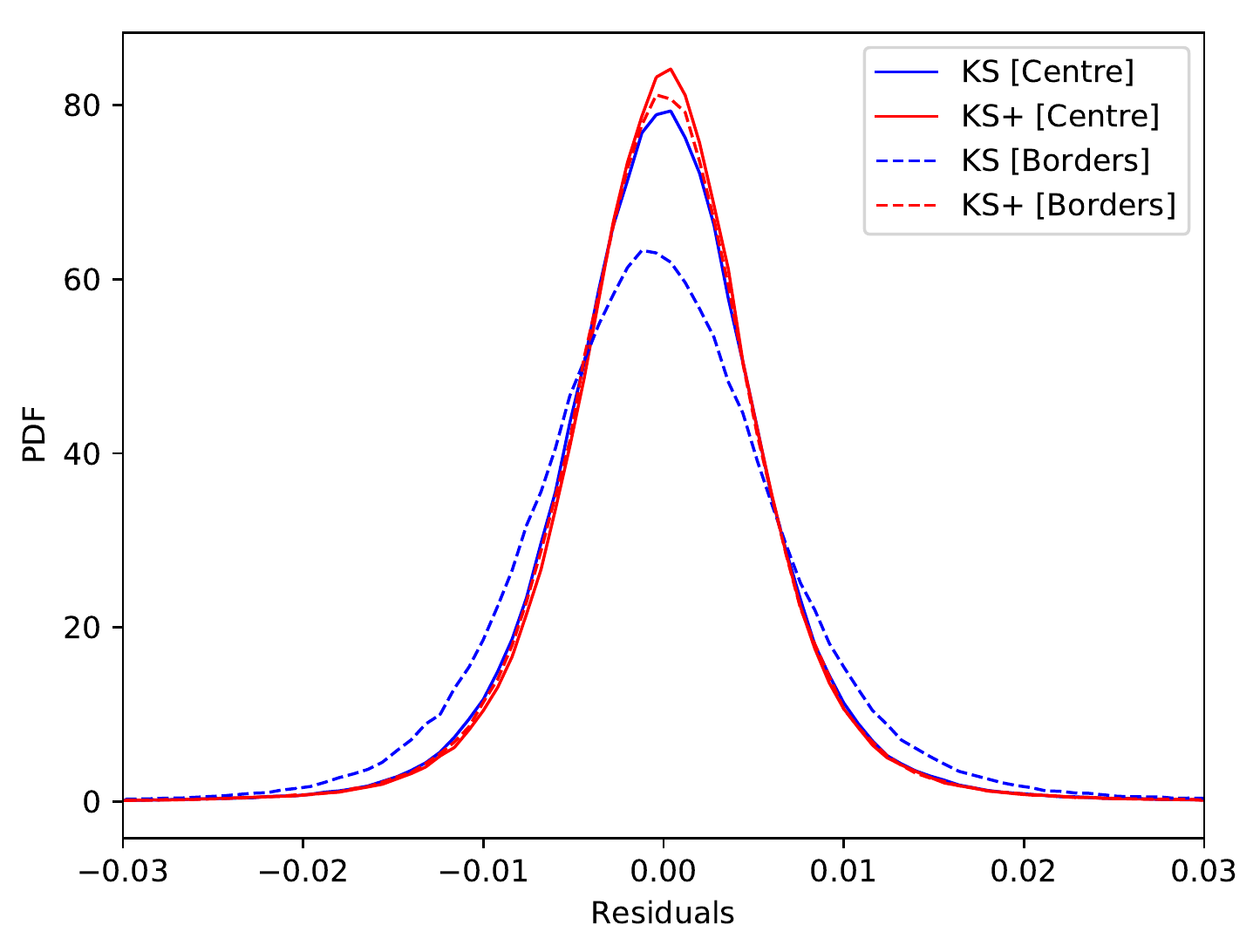,height=6.5cm,clip=}
}
\caption{Field border effects: PDF of the residual errors between the original E-mode convergence map and the convergence maps reconstructed using KS (blue) and KS+ (red). The dotted lines correspond to the PDF of the residual errors measured at the boundaries of the field, and the solid lines show the PDF of the residual errors measured in the centre of the field. The borders are 100 pixels wide.}
\label{border_residual_PDF}
\end{figure}

Fig.~\ref{kappa_borders} compares the results of the KS (left) and KS+ (right) methods for border effects. It shows the residual error maps corresponding to the pixel difference between the original E-mode convergence map and the reconstructed maps.
With KS, as expected, the pixel difference shows errors at the border of the field. With KS+, there are also some low-level boundary effects, but these errors are considerably reduced and do not show any significant structure at the field border.
In KS+, the image is extended to reduce the border effects. The effect of borders decreases when the size of the borders increases. A border size of 512 pixels has been selected for \Euclid as a good compromise between precision and computational speed. It corresponds to extending the image to be in-painted to 2048 $\times$ 2048 pixels.
Again, the PDF of these residuals can be compared to quantify the errors. For the two methods, Fig.~\ref{border_residual_PDF} shows the residuals PDFs computed at the boundaries (as dotted lines) and in the remaining central part of the image (as solid lines). The border width used to compute the residual PDF is 100 pixels, which corresponds to about one degree. 
With the KS method, the standard deviation of the residuals in the centre of the field is 0.0062.
 In the outer regions, the border effect causes errors of 0.0076 (i.e. 25\% larger than at the centre). Away from the borders, the KS+ method gives results similar to the KS method (0.0060). However, it performs much better at the border, where the error only reaches 0.0061. The small and uniform residuals of the KS+ method show how efficiently it corrects for borders effects.

\begin{figure}
\vbox{
\centerline{
\psfig{figure=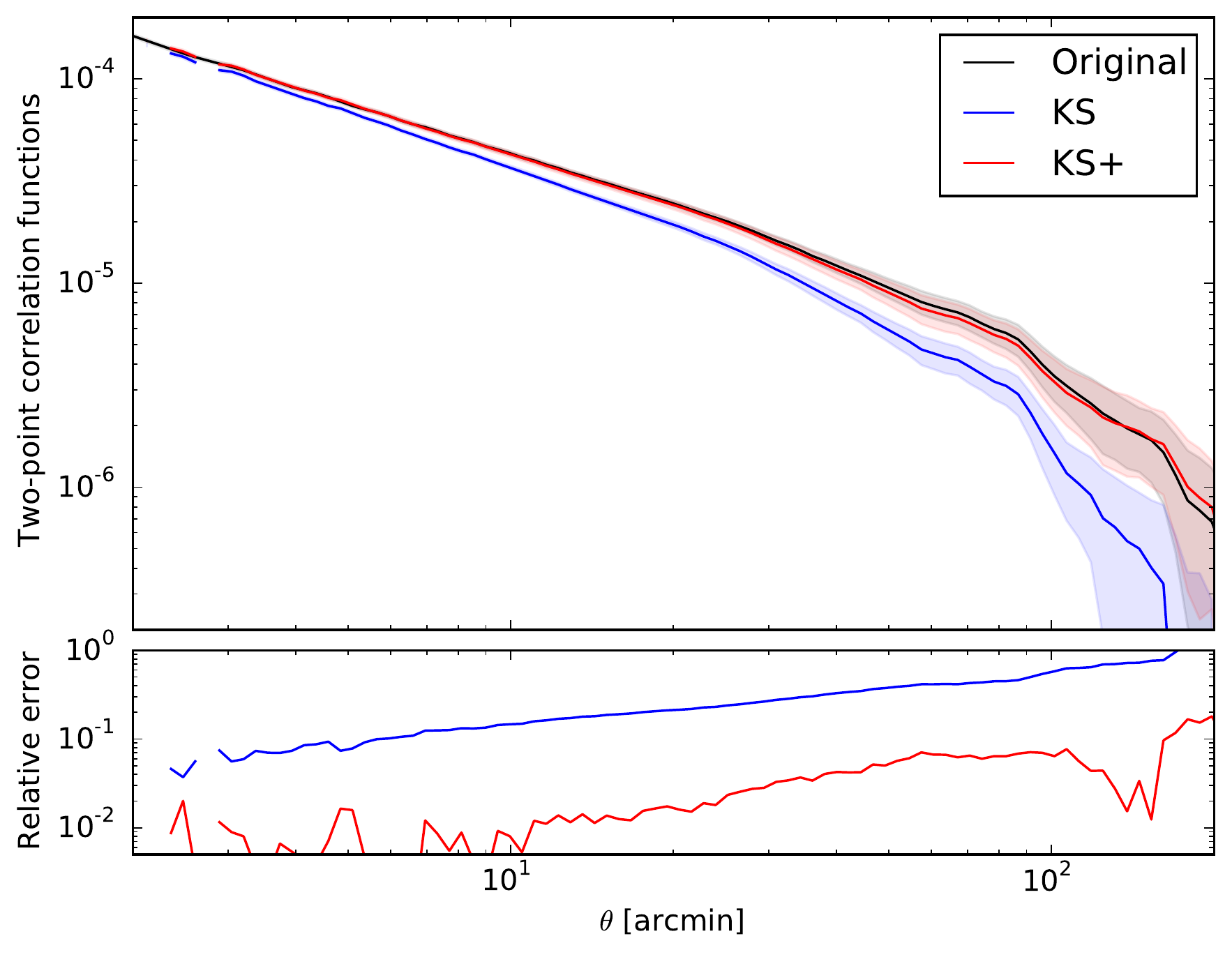,width=9.cm,clip=}
}
\caption{Field border effects: Mean shear two-point correlation function $\xi_+$ (black) compared to the corresponding mean convergence two-point correlation function $\xi_{\kappa_{\rm E}}$ reconstructed using the KS method (blue) and the KS+ method (red). The shaded area represents the uncertainties on the mean estimated on 1000 deg$^{2}$. The lower panel shows the relative two-point correlation error introduced by border effects.}
\label{border_corr}
}
\end{figure}

As before, the scale dependence of the errors can be estimated using the two-point correlation function and higher-order moments computed at different scales. 
Fig.~\ref{border_corr} shows the two-point correlation functions.
For both methods, the errors increase with angular scale because the fraction of pairs of pixels that include boundaries increase with scale. The loss of amplitude at the image border is responsible for significant errors in the two-point correlation function of the KS convergence maps. 
In contrast, the errors are about five to ten times smaller with the KS+ method and remain in the $1\sigma$ uncertainty range of the original two-point correlation function.

\begin{figure}
\vbox{
\centerline{
\hbox{
\psfig{figure=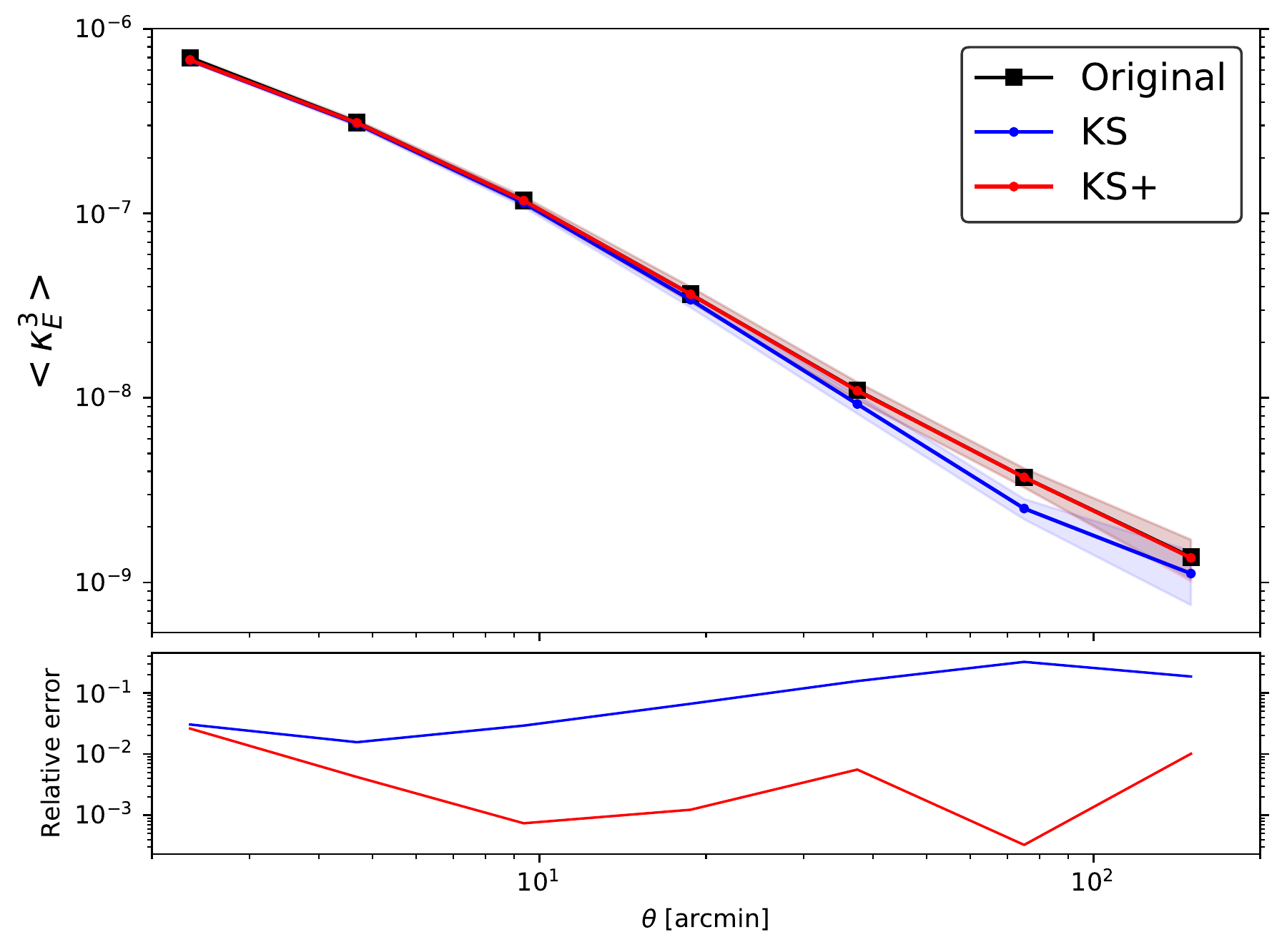,width=9.cm,clip=}
}}
\vspace{0.3cm}
\centerline{
\hbox{
\psfig{figure=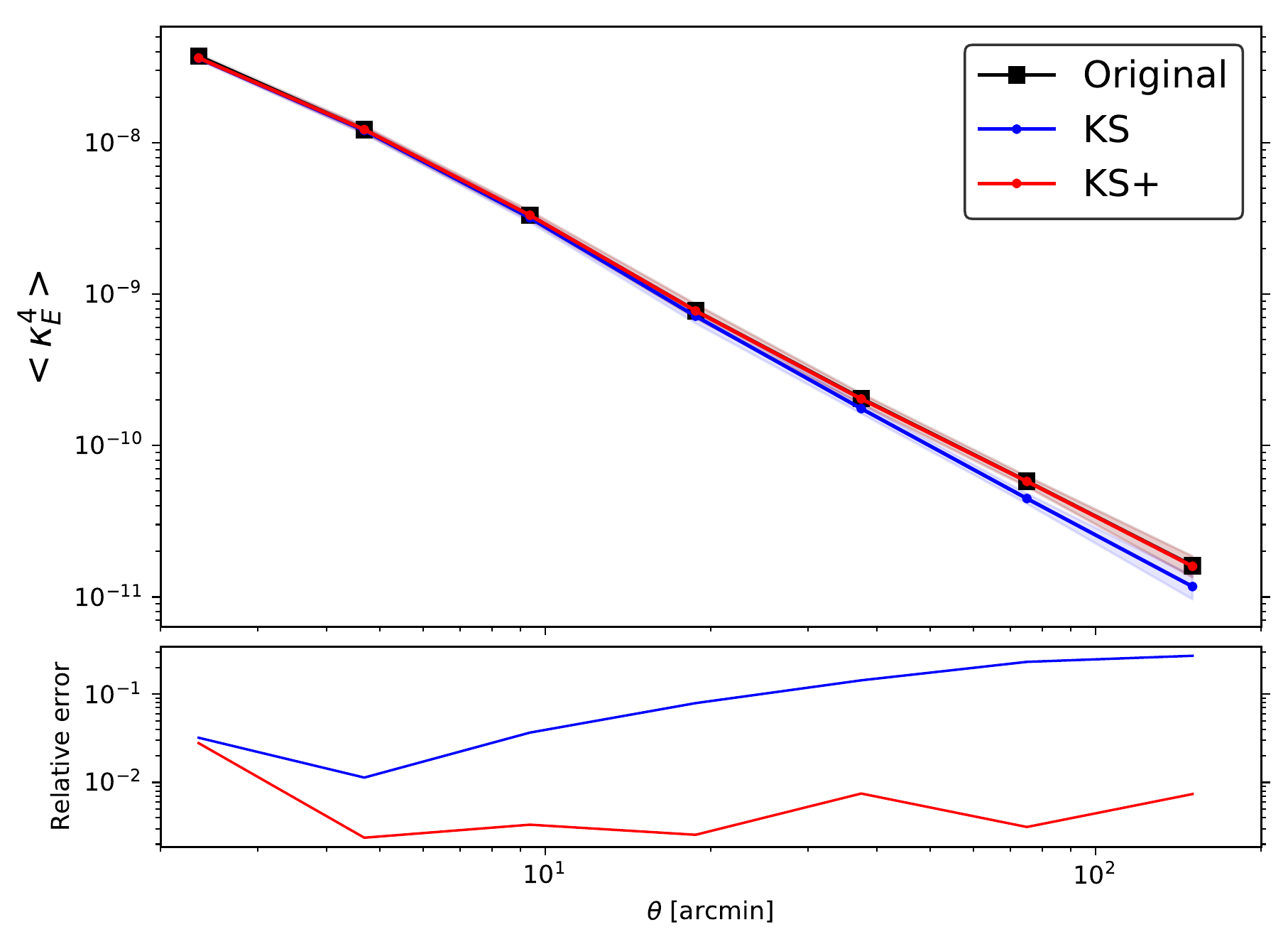,width=9.cm,clip=}
}}}
\caption{Field border effects: Third-order (upper panel) and fourth-order (lower panel) moments estimated on seven wavelet bands of the original convergence (black) compared to the moments estimated on the KS (blue) and KS+ (red) convergence maps reconstructed from noise-free shear maps. The shaded area represents the uncertainties on the mean estimated on 1000 deg$^{2}$. The lower panel shows the relative higher-order moment errors introduced by border effects.}
\label{border_hos}
\end{figure}

Fig.~\ref{border_hos} shows field borders effects on the third-order (upper panel) and fourth-order (lower panel) moments of the convergence maps at different scales.
As was observed earlier for the two-point correlation estimation, the KS method introduces errors at large scales on the third- and fourth-order moment estimation. With KS+, the discrepancy is about $1\%$ and within the $1\sigma$ uncertainty.

When the two-point correlation functions and higher-order moments are computed far from the borders, the errors of the KS method decrease, as expected. In contrast, we observe no significant improvement when the statistics are computed similarly on the KS+ maps, indicating that KS+ corrects for borders properly.

\subsection{Reduced shear}
In this section we quantify the errors due to the approximation of shear ($\gamma$) by the reduced shear ($g$).
To this end, we used the noise-free shear fields described in Sect.~\ref{sect_simu} and computed the reduced shear fields using Eq.~(\ref{reducedshear}) and the convergence provided by the catalogue. We then derived the reconstructed convergence maps using the KS and KS+ methods.

\begin{figure}
\vbox{
\centerline{
\psfig{figure=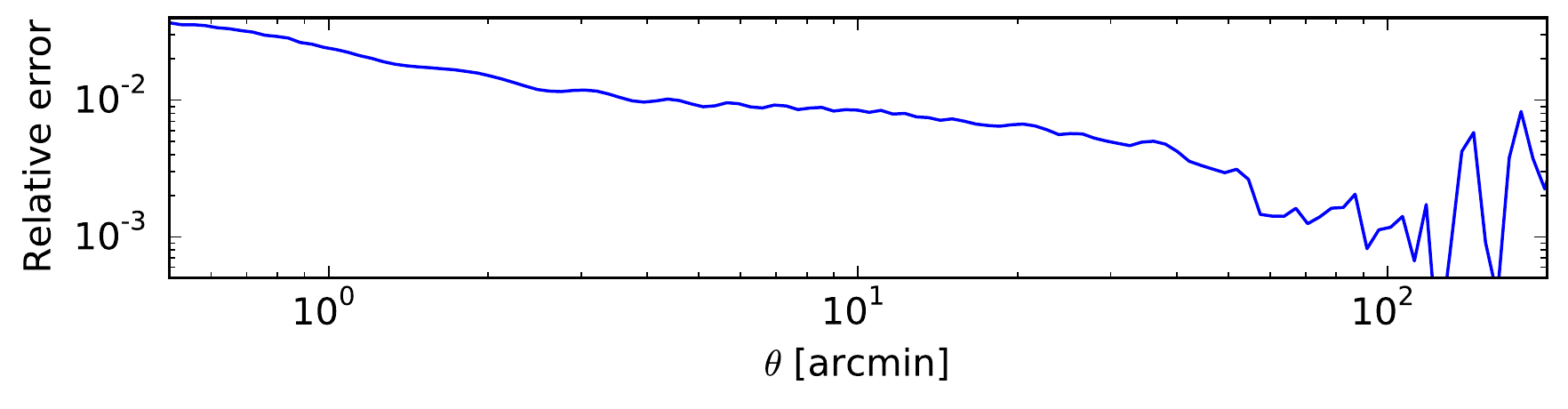,width=9.cm,clip=}
}
\caption{Reduced shear effects: Relative two-point correlation error between the mean two-point correlation functions \smash{$\xi_+^{\gamma}$} estimated from the shear fields and corresponding mean two-point correlation function \smash{$\xi_+^{\rm g}$} estimated from the reduced shear fields without any correction.}
\label{reduced_corr}
}
\end{figure}

For both methods, the errors on the convergence maps are dominated by field border effects. 
We did not find any estimator able to separate these two effects and then identify the reduced shear effect in the convergence maps. 
The errors introduced by the reduced shear can be assessed by comparing the shear and reduced shear two-point correlation functions (see Fig.~\ref{reduced_corr}), however. While the differences are negligible at large scales, they reach the percent level on arcminute scales \citep[in agreement with][]{wl:white05}, where they become comparable or larger than the KS+ errors due to border effects.

\subsection{Shape noise}

In this section we study the effect of the shape noise on convergence maps.
We derived noisy shear maps, assuming a Gaussian noise ($\sigma_{\epsilon} = 0.3$).
Then, we compared the two mass-inversion methods. 
The pixel difference cannot be used in this case because the convergence maps are noise dominated (see Fig.~\ref{kappa_noise}, upper right panel).
\begin{figure}
\vbox{
\centerline{
\psfig{figure=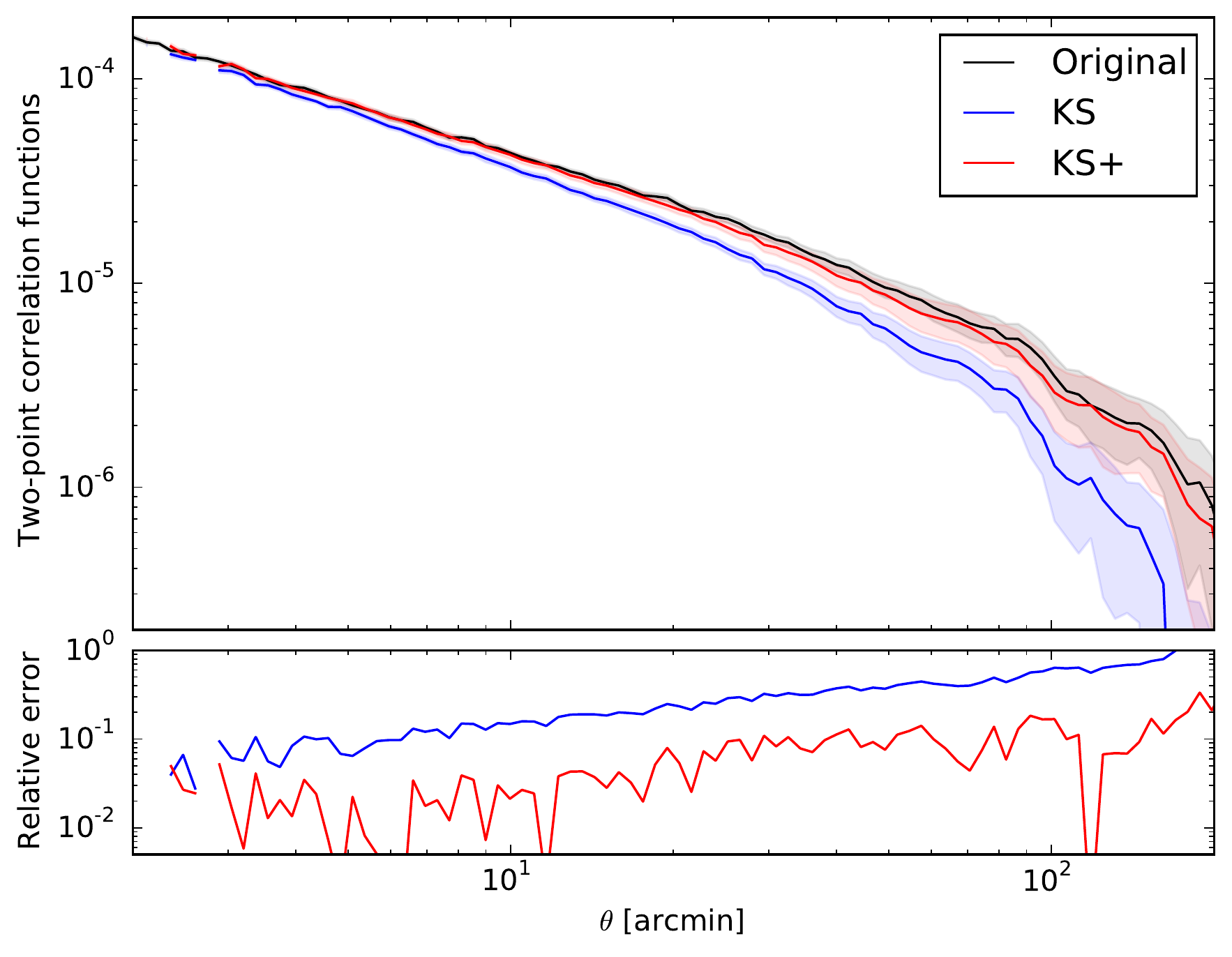,width=9.cm,clip=}
}
\caption{Shape noise effects: Mean shear two-point correlation function $\xi_+$ (black) and corresponding mean convergence two-point correlation function $\xi_{\kappa_{\rm E}}$ estimated from complete noisy shear fields. The convergence maps have been estimated using the KS method (blue) and using the KS+ method (red). The shaded area represents the uncertainties on the mean estimated on 1000 deg$^{2}$. The lower panel shows the relative two-point correlation error introduced by shape noise.}
\label{border_noise_corr}
}
\end{figure}
However, we can still assess the quality of the convergence maps using two-point correlation functions because the ellipticity correlation is an unbiased estimate of the shear correlation, and similarly, the convergence two-point correlation functions is unbiased by the shape noise.

Fig.~\ref{border_noise_corr} compares the results of the KS and KS+ methods when shape noise is included.
Compared to Fig.~\ref{border_corr}, the two-point correlation of the noisy maps is less smooth because the noise fluctuations do not completely average out. However, the amplitude of the errors introduced by the mass inversion remain remarkably similar to the errors computed without shape noise for the KS and KS+ methods. The same conclusions then hold: the errors are about five times smaller with the KS+ method.

\begin{figure}
\vbox{
\centerline{
\hbox{
\psfig{figure=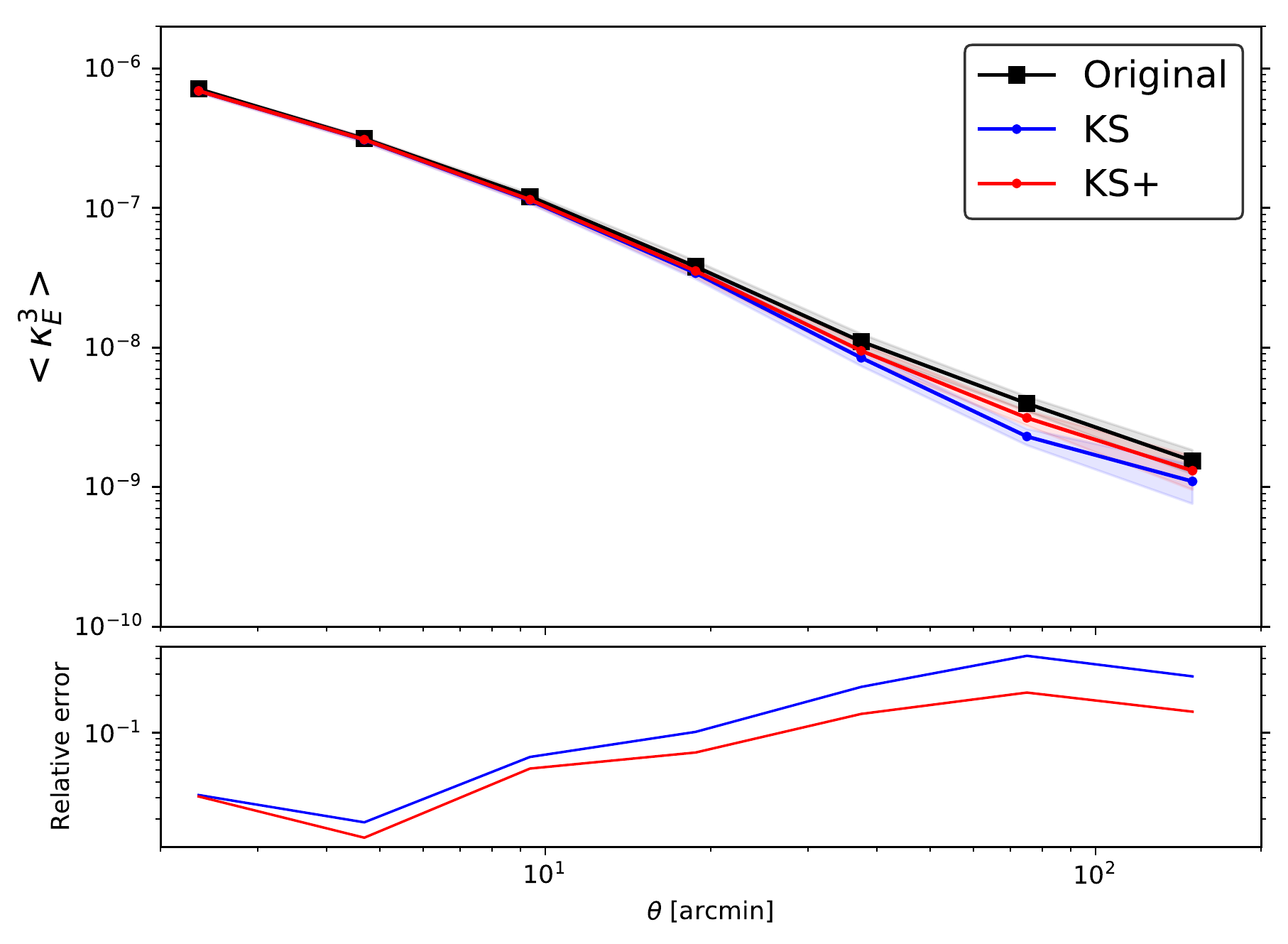,width=9.cm,clip=}
}}
\vspace{0.3cm}
\centerline{
\hbox{
\psfig{figure=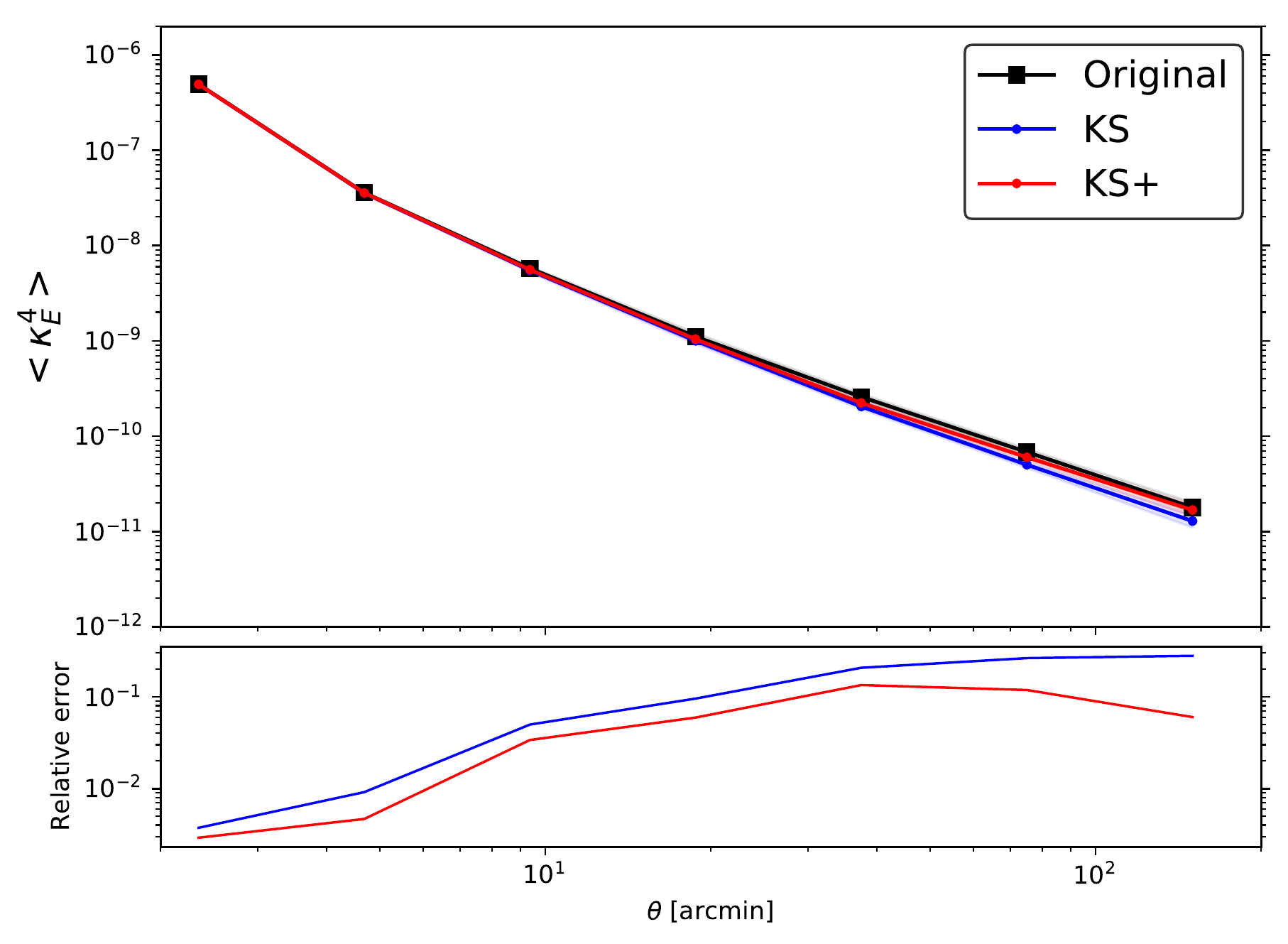,width=9.cm,clip=}
}}}
\caption{Shape noise effects: Third-order (upper panel) and fourth-order (lower panel) moments estimated on seven wavelet bands of the original convergence with realistic shape noise (black) compared to the moments estimated on the KS (blue) and KS+ (red) convergence reconstructed from noisy shear maps. The shaded area represents the uncertainties on the mean estimated on 1000 deg$^{2}$. The lower panel shows the relative higher-order moment errors introduced by shape noise.}
\label{noise_border_hos}
\end{figure}

Moments of noisy maps are biased and potentially dominated by the shape noise contribution. For instance, the total variance in the noisy convergence map is expected to be the sum of the variance in the noise-free convergence map and the noise variance. Therefore moments of the noisy KS and KS+ convergence maps cannot be directly compared to moments of the original noise-free convergence maps. 
Instead, Fig.~\ref{noise_border_hos} compares them to the moments of the original convergence maps where noise was added with properties similar to the noise expected in the convergence maps. For this purpose, we generated noise maps $N_1$ and $N_2$ for each field using Eq.~(\ref{eq:noise1}) and (\ref{eq:noise2}), and we derived the noise to be added in the convergence using Eq.~(\ref{eq:noise3}).

The comparison of Fig.~\ref{noise_border_hos} to Fig.~\ref{border_hos} shows that the third-order moment of the convergence is not affected by shape noise. In contrast, the fourth-order moment is biased for scales smaller than 10$\arcmin$. The two methods slightly underestimate the third- and fourth-order moments at large scales. However, with KS+, the errors are reduced by a factor of 2 and remain roughly within the $1\sigma$ uncertainty.

\subsection{All systematic effects taken into account simultaneously}
In this section, we assess the performance of KS and KS+ for realistic data sets by combining the effects of shape noise, reduced shear, borders, and missing data.

\begin{figure*}
\vbox{
\centerline{
\hbox{
\psfig{figure=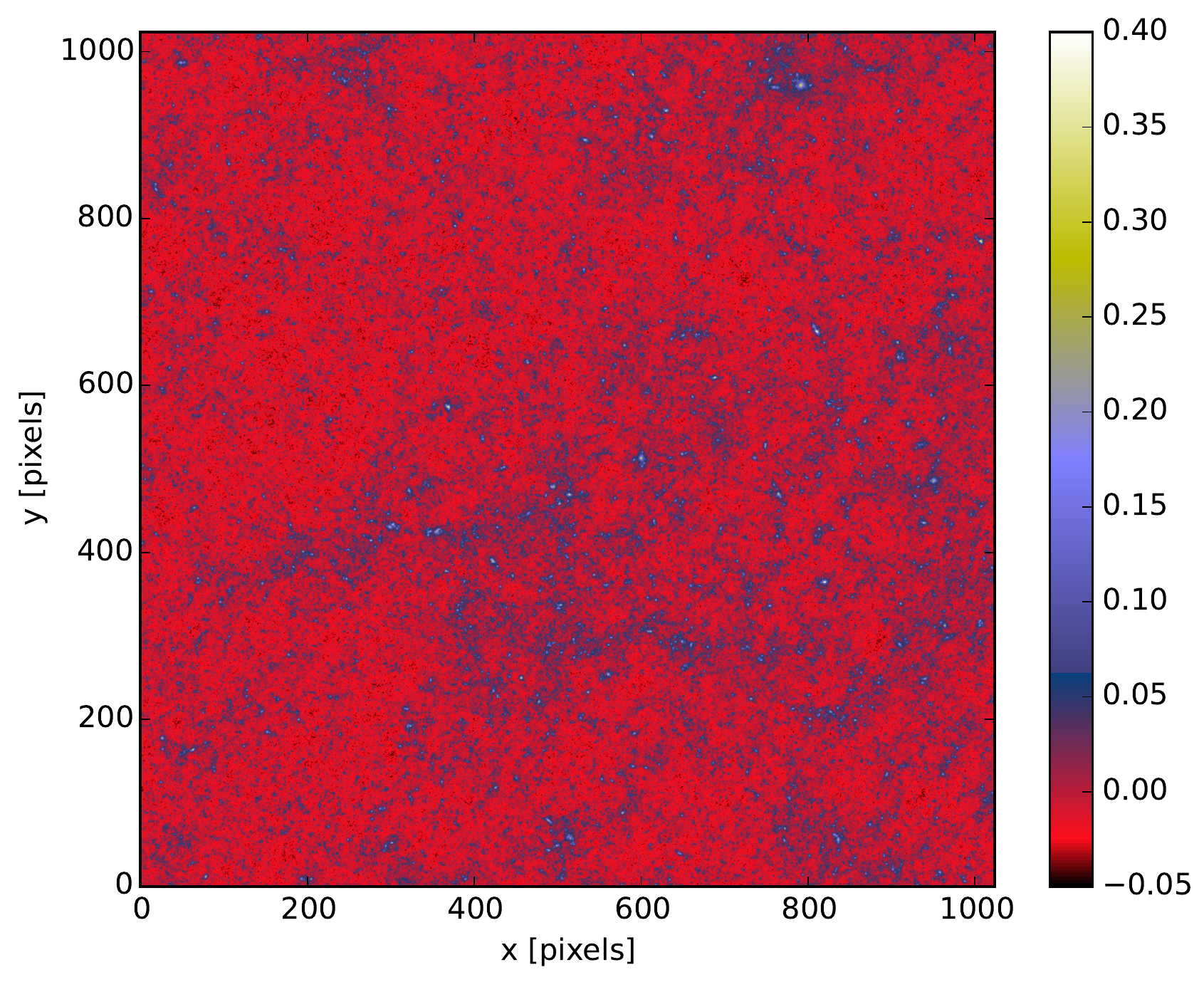,height=6.cm,width=6.7cm,clip=}
\hspace{0.6cm}
\psfig{figure=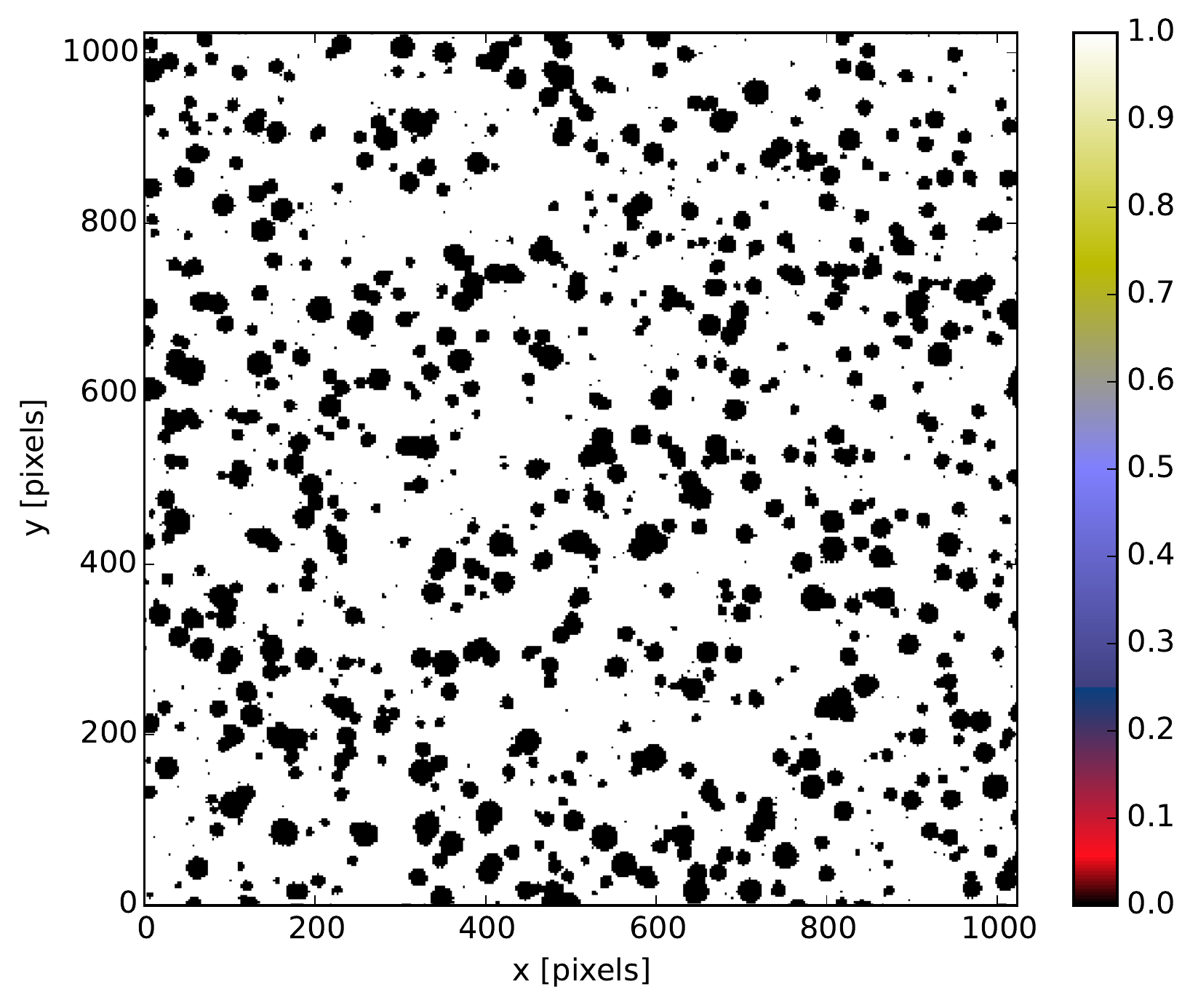,height=6.cm,width=6.4cm,clip=}
}}
\vspace{0.3cm}
\centerline{
\hbox{
\psfig{figure=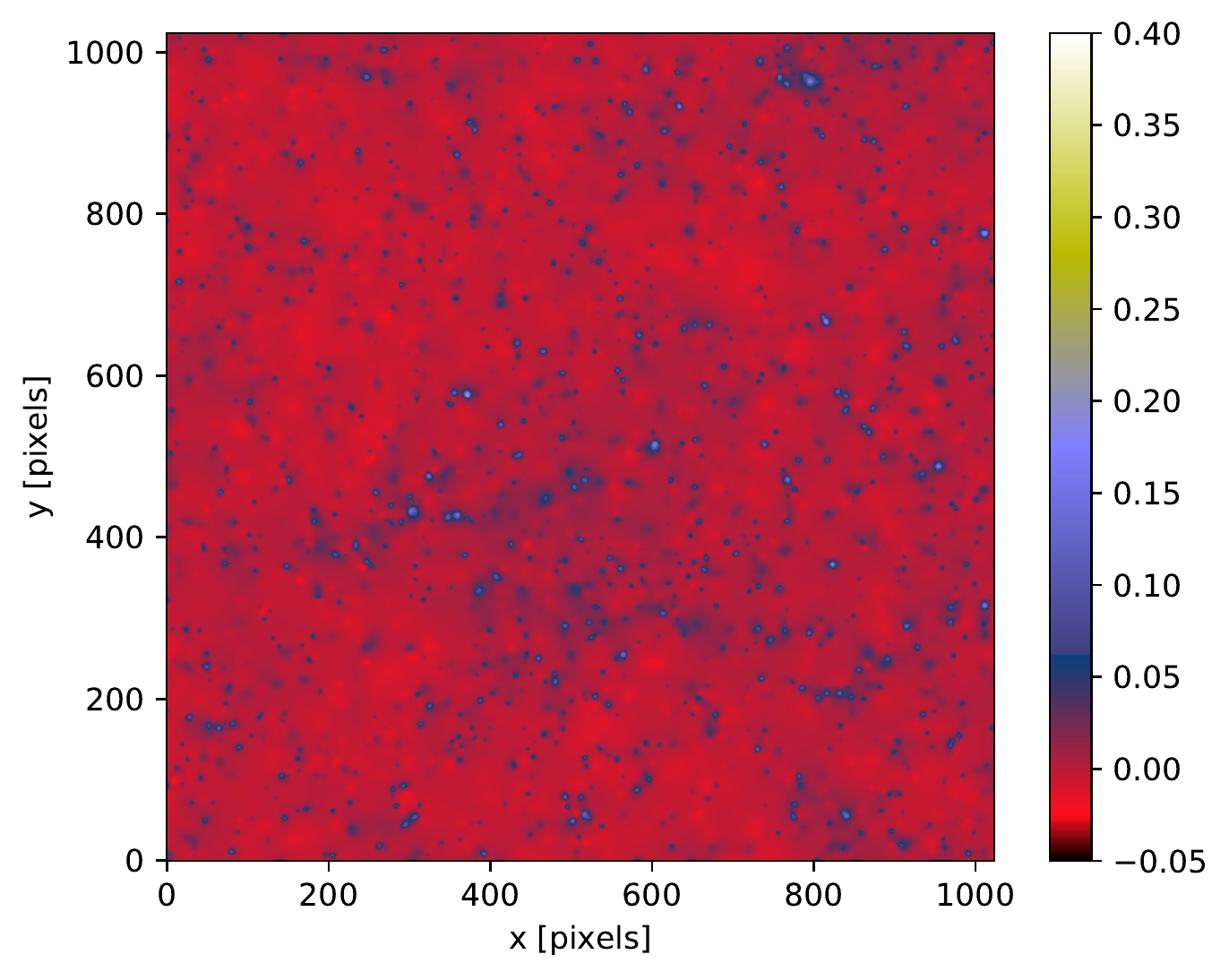,height=6.cm,width=7cm,clip=}
\hspace{0.2cm}
\psfig{figure=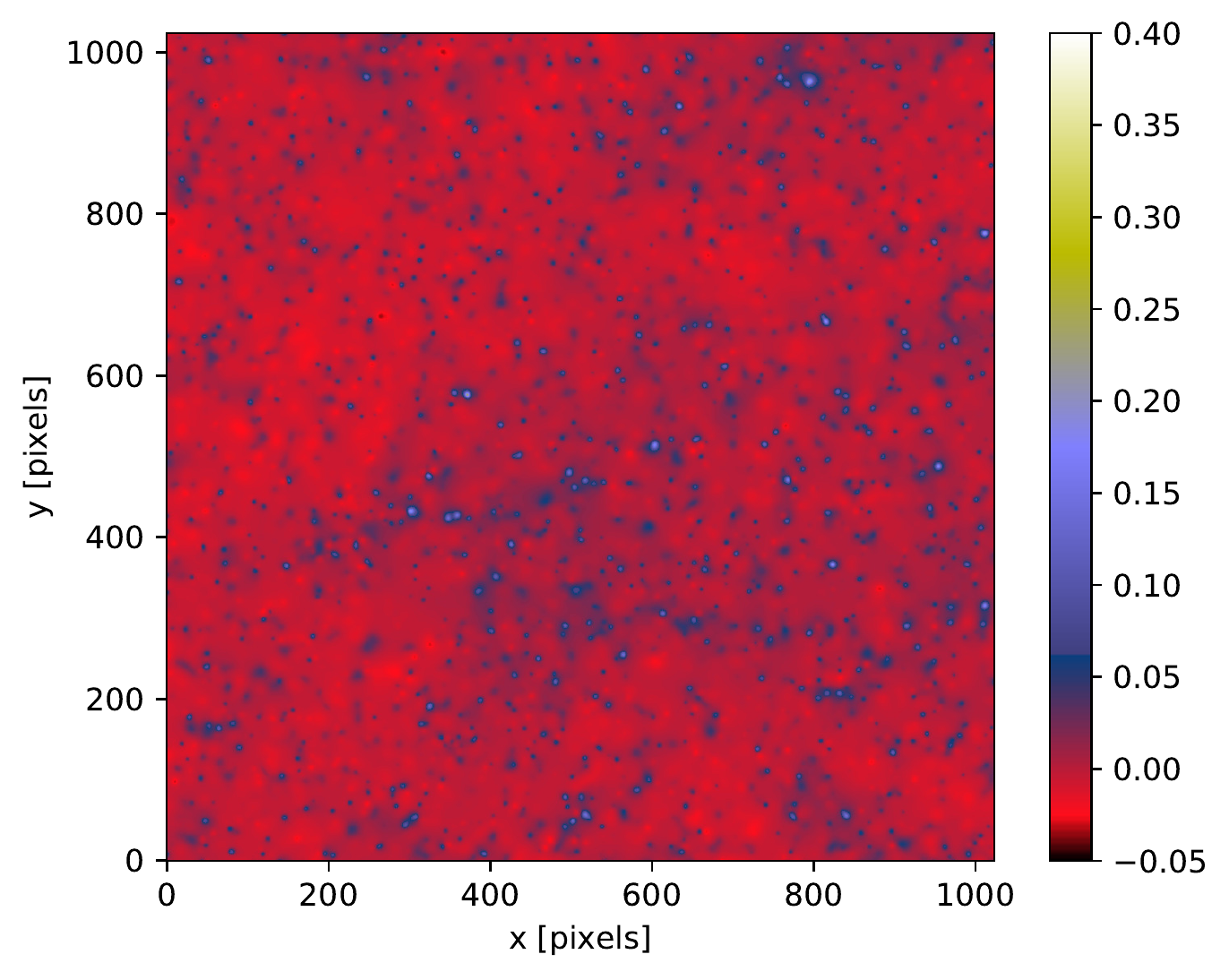,height=6.cm,width=7cm,clip=}
}}

}
\caption{All systematic effects: The upper panels show the original E-mode convergence $\kappa$ map (left) and the mask that is applied to the shear maps (right). The lower panels show the convergence map reconstructed from an incomplete noisy shear field using the KS method (left) and using the KS+ method (right) applying a nonlinear MRLens filtering with $\alpha_{\rm FDR} = 0.05$. The field is $10^\circ \times 10^\circ$ downsampled to $1024 \times 1024$ pixels.}

\label{kappa_complete}
\end{figure*}

Fig.~\ref{kappa_complete} compares the results of the KS method and the KS+ method combined with a filtering step to correct for all systematic effects in one field.
We used the nonlinear MRLens filter to reduce the noise in the KS and KS+ convergence maps because it is particularly well suited for the detection of isotropic structures \citep{stat:pires09a, wl:pires12, wl:lin16}. Again, KS+ better recovers the over-densities because it reduces the signal leakage during the mass inversion compared to KS.

\begin{figure}
\vbox{
\centerline{
\psfig{figure=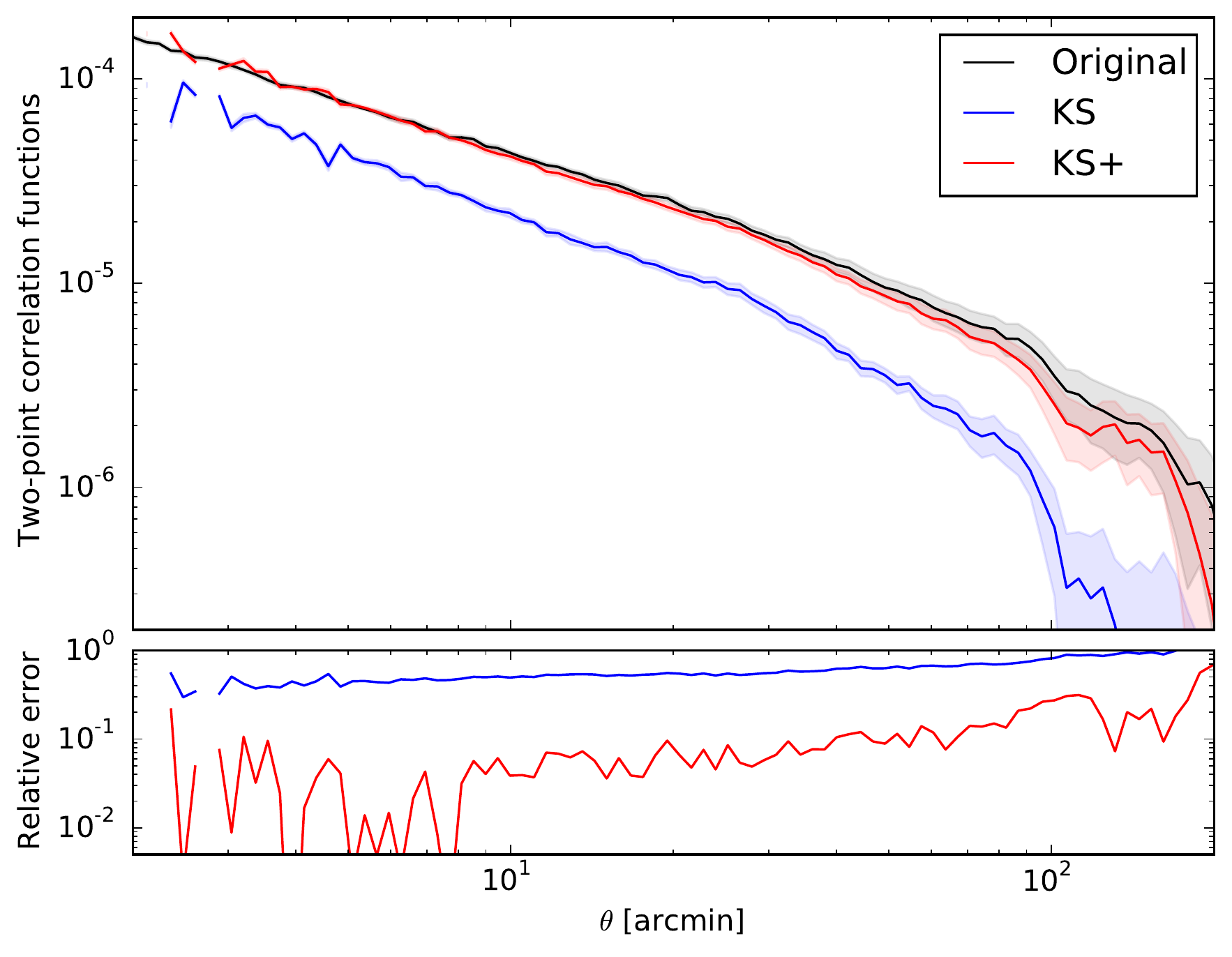,width=9.cm,clip=}
}
\caption{All systematic effects: Mean shear two-point correlation function $\xi_+$ (black) and corresponding mean convergence two-point correlation function $\xi_{\kappa_{\rm E}}$ estimated from incomplete noisy shear fields. The convergence maps have been estimated using KS (blue) and KS+ (red). The convergence two-point correlations were estimated outside the mask. The shaded area represents the uncertainties on the mean estimated on 1000 deg$^{2}$. The lower panel shows the normalised difference between the two upper curves.}
\label{missing_noise_corr} 
}
\end{figure}

Fig.~\ref{missing_noise_corr} shows the two-point correlation computed with the two methods. 
The masked regions were excluded from the two-point correlation computation, resulting in fewer pairs and higher noise than in Fig.~\ref{border_noise_corr}.
Again, the strong leakage due to missing data is clearly observed with the KS method. 
The results obtained with the KS+ method reduce the errors in the mean convergence two-point correlation function by a factor of about 5, and the errors remain roughly within the 1$\sigma$ uncertainty.

\begin{figure}
\vbox{
\centerline{
\hbox{
\psfig{figure=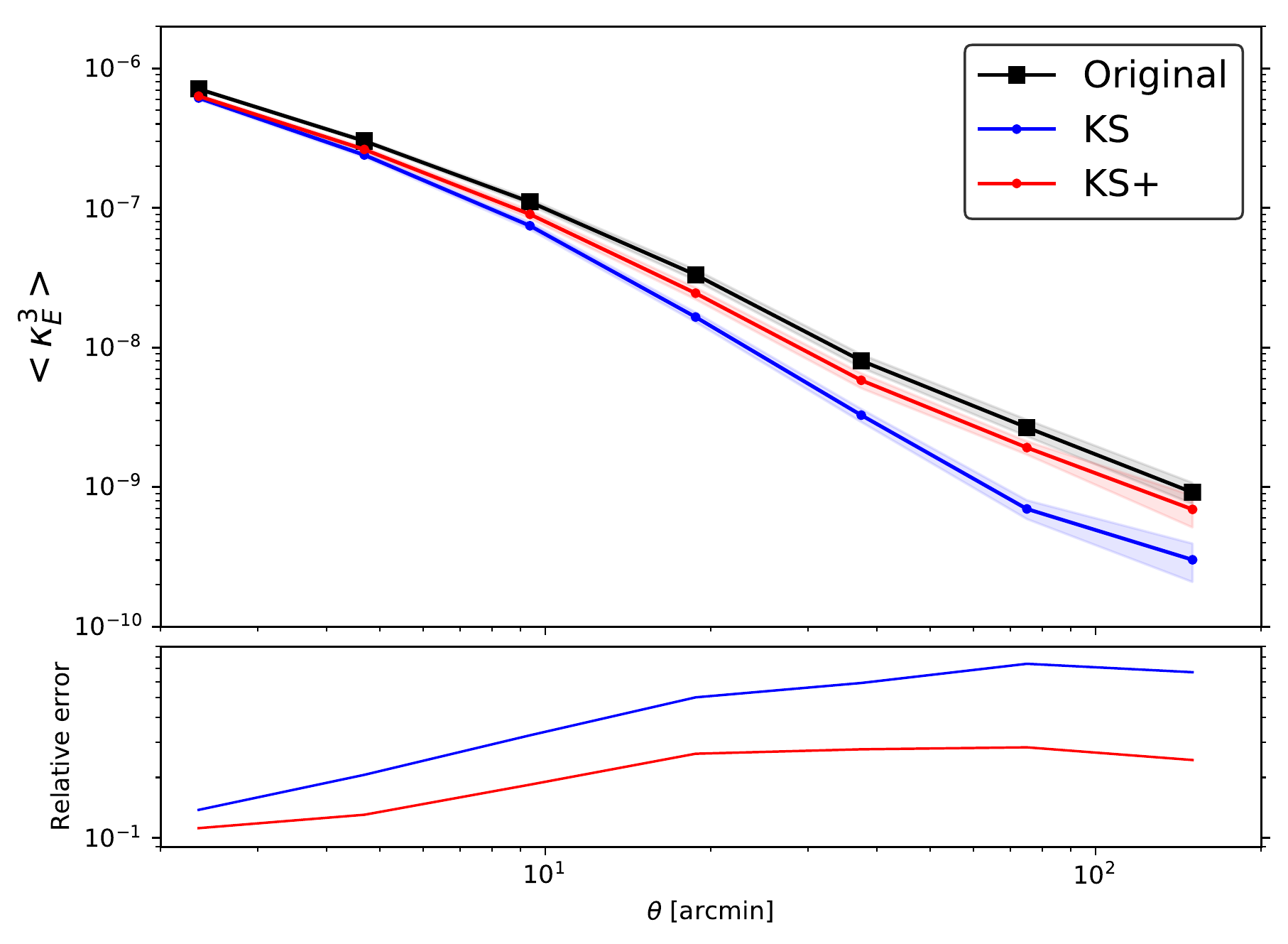,width=9.cm,clip=}
}}
\vspace{0.3cm}
\centerline{
\hbox{
\psfig{figure=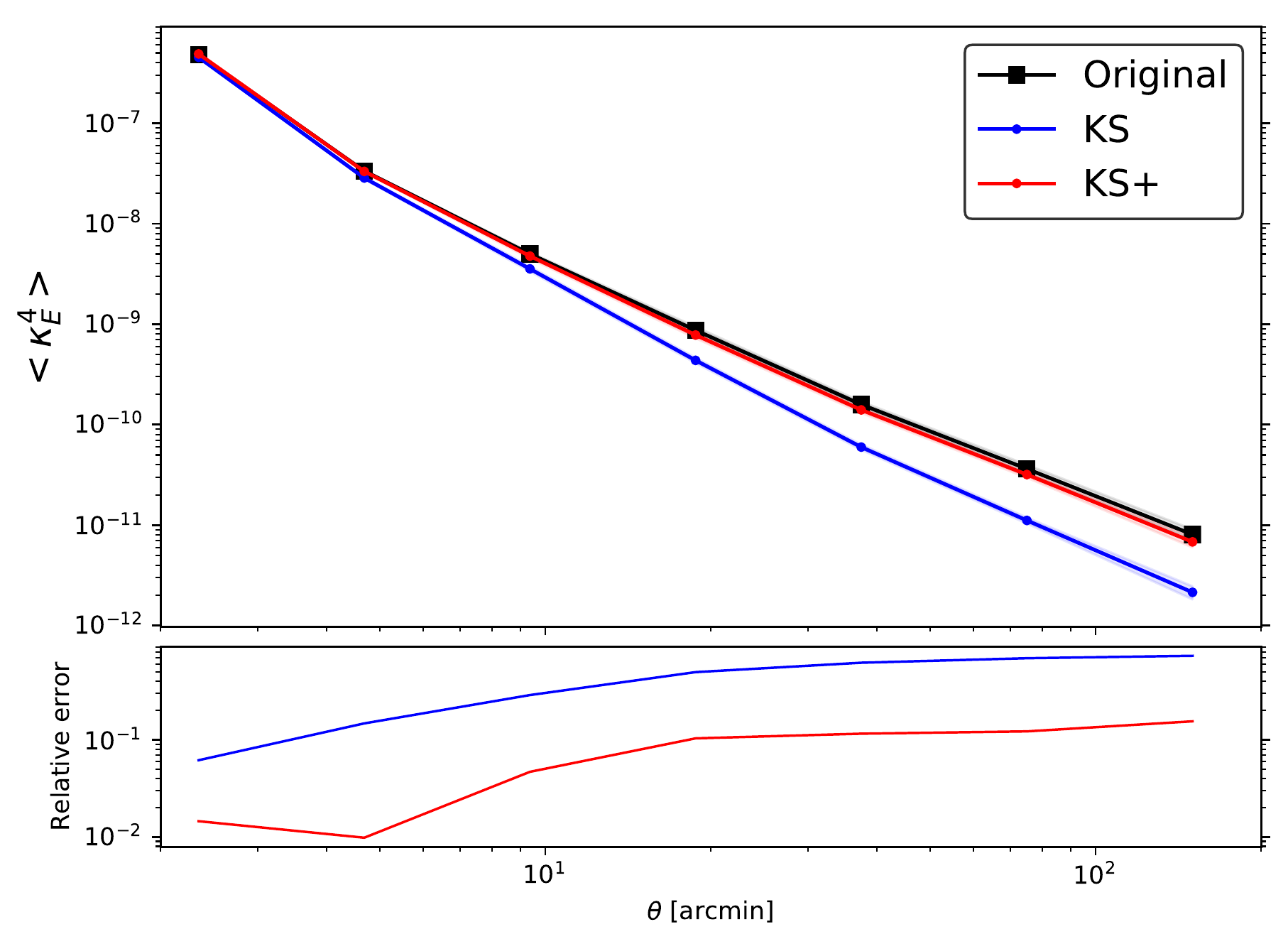,width=9.cm,clip=}
}}}
\caption{All systematic effects: Third-order (upper panel) and fourth-order (lower panel) moments estimated on seven wavelet bands of the original convergence with realistic noise (black) compared to the moments estimated using KS (blue) and KS+ (red) obtained from incomplete noisy shear maps. The third- and fourth-order moments are estimated outside the mask. The shaded area represents the uncertainties on the mean estimated on 1000 deg$^{2}$. The lower panel shows the relative higher-order moment errors.}
\label{noise_missing_hos}
\end{figure}

In Fig.~\ref{noise_missing_hos} we test the efficacy of the mass-inversion methods in preserving higher-order moments of the convergence maps in a realistic setting. As before, realistic noise was added to the original convergence maps for comparison.
As was observed earlier in the noise-free case, the KS method systematically underestimates the third- and fourth-order moments at all scales. 
With KS+, the errors are significantly reduced, by a factor of about 2 in the third-order moment and by a factor of about 10 in the fourth-order moment estimation, at all scales.
Although reduced, the errors of the KS+ method on the third-order moment cannot be neglected.
These errors might result from noise correlations introduced by the inpainting method in the shear maps. Inside the gaps, the noise is indeed correlated because it is interpolated from the remaining data. These noise correlations propagate into the convergence maps and can explain the bias in the moment estimation.

We note that the two-point correlation functions and higher-order moments are here only used to probe the accuracy of the reconstruction methods. For specific applications, the small residuals of the KS+ method can be reduced even more using additional treatment such as down-weighting the region around the mask when the moments are computed \citep[e.g.][]{cfhtlens:vanwaerbeke13}.

\section{Conclusion}
\label{sect_cl}

This paper was motivated by the use of convergence maps in \Euclid to constrain cosmological parameters and to assess other physical constraints. 
Convergence maps encode the lensing information in a different manner, allowing more optimised computations than shear.
However, the mass-inversion process is subject to field border effects, missing data, reduced shear, intrinsic alignments, and shape noise. This requires accurate control of the systematic effects during the mass inversion to reduce the information loss as much as possible. 
We presented and compared the two mass-inversion methods that are included in the official Euclid data-processing pipeline: the standard Kaiser \& Squires (KS) method, and an improved Kaiser \& Squires (KS+) mass-inversion technique that integrates corrections for the mass-mapping systematic effects.
 The systematic effects on the reconstructed convergence maps were studied using the Euclid Flagship mock galaxy catalogue.

In a first step, we analysed and quantified one by one the systematic effects on reconstructed convergence maps using two-point correlation functions and moments of the convergence.
In this manner, we quantified the contribution of each effect to the error budget to better understand the error distribution in the convergence maps. With KS, missing data are the dominant effect at all scales. Field border effects also have a strong effect, but only at the map borders. These two effects are significantly reduced with KS+. The reduced shear is the smallest effect in terms of contribution and only affects small angular scales. 
The study also showed that pixellisation provides an intrinsic regularisation and that no additional smoothing step is required to avoid infinite noise in the convergence maps.

In a second step, we quantified the errors introduced by the KS and KS+ methods in a realistic setting that included the systematic effects. 
We showed that the KS+ method reduces the errors on the two-point correlation functions and on the moments of the convergence compared to the KS method. 
The errors introduced by the mass inversion on the two-point correlation of the convergence maps are reduced by a factor of about 5. 
The errors on the third-order and fourth-order moment estimates are reduced by factors of about 2 and 10, respectively.
Some errors remain in the third-order moment that remain within the $2\sigma$ uncertainty. They might result from noise correlations introduced by the inpainting method inside the gaps.

Our study was conducted on a mock of 1000 deg$^2$ divided into ten fields of 10$^{\circ}$ $\times$ 10$^{\circ}$ to remain in the flat-sky approximation. \Euclid will observe a field of 15 000 deg$^2$. As long as KS+ has not been extended to the curved sky, it is not possible to apply the method to larger fields without introducing significant projection effects. However, the \Euclid survey can be divided into small fields, which allows reducing the uncertainties in the statistics that are estimated on the convergence maps. Moreover, we can expect that part of the errors will average out.

Recent studies have shown that combining the shear two-point statistics with higher-order statistics of the convergence such as higher-order moments \citep{hos:vicinanza18}, Minkowski functionals \citep{hos:vicinanza19}, or peak counts \citep{hos:liu15,hos:martinet18} allows breaking common degeneracies. 
The precision of the KS+ mass inversion makes the E-mode convergence maps a promising tool for such cosmological studies.
In future work, we plan to propagate these errors into cosmological parameter constraints using higher-order moments and peak counts.

\label{sect_cl}

\section*{Acknowledgments}
{This study has been carried inside the Mass Mapping Work Package of the Weak Lensing Science Working Group of the \Euclid project to better understand the impact of the mass inversion systematic effects on the convergence maps. 
The authors would like to thank the referees and editors for their valuable comments, which helped to improve the manuscript.
S. Pires thanks F. Sureau, J. Bobin, M. Kilbinger, A. Peel and J.-L. Starck for useful discussions. 
\AckEC

\section*{Appendix A: KS+ inpainting algorithm}

This appendix describes the KS+ method presented in Sect.~\ref{iks} in more detail.
The solution of the KS+ mass inversion is obtained through the iterative algorithm described in Algorithm~\ref{algo}.

The outer loop starting at step 5 is used to correct for the reduced shear using the iterative scheme described in Sect.~\ref{reduced}. The inner loop starting at step 7 is used to solve the optimisation problem defined by Eq.~(\ref{eq2}). 
$\bm \Phi$ is the discrete cosine transform operator matrix.
If the convergence ${\bm \kappa}$ is sparse in $\bm \Phi$, most of the signal is contained in the strongest DCT coefficients. The smallest coefficients result from missing data, border effects, and shape noise. Thus, the algorithm is based on an iterative algorithm with a threshold that decreases exponentially (at each iteration) from a maximum value to zero, following the decreasing law $F$ described in \cite{wl:pires09}. 
By accumulating increasingly more high DCT coefficients through each iteration, the gaps in $\bm{\tilde \gamma}$ fill up steadily, and the power of the spurious B modes due to the gaps decreases.
The algorithm uses the fast Fourier transform at each iteration to compute the shear maps $\bm \gamma$ from the convergence maps $\bm \kappa$ (step 14) and the inverse relation (step 16).

A data-driven power spectrum prior is introduced at steps 11-13. To do so, the KS+ algorithm uses the undecimated isotropic wavelet transform that decomposes an image $\bm \kappa$ into a set of coefficients $\{ \bm{w_1}, \bm{w_2}, ..., \bm{w_J}, \bm{c_J}\}$, as a superposition of the form
\begin{eqnarray}
\bm \kappa[i_1, i_2]= \bm{c_{J}}[i_1, i_2] + \sum_{j=1}^{J} \bm{w_{j}}[i_1,i_2], 
  \label{wavelet}
 \end{eqnarray}
where $\bm{c_{J}}$ is a smoothed version of the image, and $\bm \kappa$ and $\bm{w_{j}}$ are a set of aperture mass maps (usually called wavelet bands) at scale $\theta = 2^{j}$. Then, we estimate the variance on each wavelet band $\bm{w_j}$. The variance per scale estimated in this way can be directly compared to the power spectrum. This provides a way to estimate a broadband power spectrum of the convergence $\bm \kappa$ from incomplete data. 
The power spectrum is then enforced by multiplying each wavelet coefficient by the factor \smash{$\sigma_j^{\rm{out}}/\sigma_j^{\rm{in}}$} inside the gaps, where \smash{$\sigma_j^{\rm{in}}$} and \smash{$\sigma_j^{\rm{out}}$} are the standard deviation estimated in the wavelet band $\bm{w_j}$ inside and outside the mask, respectively.
This normalisation can be described by a linear operator $\mathbf Q$ as used in Eq.~(\ref{eq2}). 
The constraint is applied on the E- and B-mode components before reconstructing the convergence $\bm \kappa$ by backward wavelet transform.
\begin{algorithm}[H]
\caption{KS+ algorithm}     
\label{algo}  
\begin{enumerate}
\item[1.] Project the shear from the celestial sphere onto a tangent plane by projecting the galaxy positions and applying a local rotation to the shear field.
\item[2.] Bin the projected shear onto a grid and define $\bm{\tilde{\gamma}}$ as the average shear in each pixel.
\item[3.] Set the mask $\mathbf M$: $M[i_1, i_2] = 1$ for pixels where we have information and $M[i_1, i_2] = 0$ for pixels with no galaxies, and take a support twice larger for the shear maps and include the borders in the masked region (see Fig.~\ref{shear_mask}).
\item[4.] Set the maximum number of iterations to $I_{\rm max}=100$, the maximum threshold $\lambda_{\rm max} = \max(\mid \bm \Phi^{\rm T} \mathbf P^* \bm{\tilde{\gamma}} \mid),$ and the minimum threshold $\lambda_{\rm min} = 0$.
\item[5.] Set $k = 0$, $\bm{\kappa_{\rm E}^{k}}=0$ and iterate:
\begin{enumerate}
\item[6.] Update the shear  $\bm{\tilde{\gamma}^{k}}  = \bm{\tilde{\gamma}} \, (1-\bm{\kappa_{\rm E}^{k}})$ and initialise the solution to $\bm{\kappa^{k}} = \mathbf P^*  \bm{\tilde{\gamma}^{k}}$.
\item[7.] Set $i = 0$, $\lambda^{0}=\lambda_{\rm max}$, $\bm{\kappa^{i}}=\bm{\kappa^{k}}$ and iterate: 
   \begin{enumerate}
    \item[8.] Compute the forward transform: $\bm \alpha = \bm{\Phi^{\rm T}} \bm{\kappa^{i}}$.
    \item[9.] Compute $\bm{\tilde \alpha}$ by setting to zero the coefficients $\bm \alpha$ below the threshold $\lambda^{i}$.
    \item[10.] Reconstruct $\bm{\kappa^{i}}$ from $\bm{\tilde \alpha}$: $\bm{\kappa^{i}} = \bm{\Phi} \bm{ \tilde\alpha}$.
    \item[11.] Decompose $\bm{\kappa^{i}}$ into its wavelet coefficients $\{ \bm{w_1}, \bm{w_2}, ..., \bm{w_J}, \bm{c_J}\}$.
    \item[12.] Renormalise the wavelet coefficients $\bm{w_j}$ by a factor $\sigma_j^{\rm{out}}/\sigma_j^{\rm{in}}$ inside the gaps.
    \item[13.] Reconstruct  $\bm{\kappa^{i}}$ by performing the backward wavelet transform from the normalised coefficients. 
    \item[14.] Perform the inverse mass relation: $\bm{\gamma^{i}} = \mathbf P  \bm{\kappa^{i}}$.
    \item[15.] Enforce the observed shear $\bm{\tilde\gamma}$ outside the gaps:
     $\bm{\gamma^{i}} = (1-\mathbf M) \, \bm{\gamma^{i}} + \mathbf M \bm{\tilde{ \gamma}^{k}}$.
     \item[16.] Perform the direct mass inversion: $\bm{\kappa^{i}} = \mathbf P^* \bm{\gamma^{i}}$.
     \item[17.] Update the threshold: $\lambda^i =  F(i, \rm \lambda_{min}, \lambda_{max})$.
     \item[18.] Set $i=i+1$. If $i<I_{\rm max}$, return to step 8.
        \end{enumerate}
 \item[19.] Set $k=k+1$, $\kappa^{k}=\kappa^{i}$. If $k < 3$, return to step 6.
\end{enumerate}
\end{enumerate}
\end{algorithm}

\section*{Appendix B: Institutions}
$^{1}$ Universit\'e Paris Diderot, AIM, Sorbonne Paris Cit\'e, CEA, CNRS F-91191 Gif-sur-Yvette Cedex, France\\
$^{2}$ Universit\"ats-Sternwarte M\"unchen, Fakult\"at f\"ur Physik, Ludwig-Maximilians-Universit\"at M\"unchen, Scheinerstrasse 1, 81679 M\"unchen, Germany\\
$^{3}$ Max Planck Institute for Extraterrestrial Physics, Giessenbachstr. 1, D-85748 Garching, Germany\\
$^{4}$ INAF-Osservatorio Astrofisico di Torino, Via Osservatorio 20, I-10025 Pino Torinese (TO), Italy\\
$^{5}$ INAF-Osservatorio Astrofisico di Torino, Via Osservatorio 20, I-10025 Pino Torinese (TO), Italy\\
$^{6}$ APC, AstroParticule et Cosmologie, Universit\'e Paris Diderot, CNRS/IN2P3, CEA/lrfu, Observatoire de Paris, Sorbonne Paris Cit\'e, 10 rue Alice Domon et L\'eonie Duquet, 75205, Paris Cedex 13, France\\
$^{7}$ Instituto de Astrof\'isica e Ci\^encias do Espa\c{c}o, Universidade do Porto, CAUP, Rua das Estrelas, PT4150-762 Porto, Portugal\\
$^{8}$ Institut de F\'isica d'Altes Energies IFAE, 08193 Bellaterra, Barcelona, Spain\\
$^{9}$ INAF-Osservatorio Astronomico di Roma, Via Frascati 33, I-00078 Monteporzio Catone, Italy\\
$^{10}$ Department of Physics "E. Pancini", University Federico II, Via Cinthia 6, I-80126, Napoli, Italy\\
$^{11}$ INFN section of Naples, Via Cinthia 6, I-80126, Napoli, Italy\\
$^{12}$ INAF-Osservatorio Astronomico di Capodimonte, Via Moiariello 16, I-80131 Napoli, Italy\\
$^{13}$ Centre National d'Etudes Spatiales, Toulouse, France\\
$^{14}$ Institute for Astronomy, University of Edinburgh, Royal Observatory, Blackford Hill, Edinburgh EH9 3HJ, UK\\
$^{15}$ ESAC/ESA, Camino Bajo del Castillo, s/n., Urb. Villafranca del Castillo, 28692 Villanueva de la Ca\~nada, Madrid, Spain\\
$^{16}$ Department of Astronomy, University of Geneva, ch. d'\'Ecogia 16, CH-1290 Versoix, Switzerland\\
$^{17}$ Institute of Space Sciences (ICE, CSIC), Campus UAB, Carrer de Can Magrans, s/n, 08193 Barcelona, Spain\\
$^{18}$ Institut d'Estudis Espacials de Catalunya (IEEC), 08034 Barcelona, Spain\\
$^{19}$ INAF-Osservatorio Astronomico di Trieste, Via G. B. Tiepolo 11, I-34131 Trieste, Italy\\
$^{20}$ INAF-Osservatorio di Astrofisica e Scienza dello Spazio di Bologna, Via Piero Gobetti 93/3, I-40129 Bologna, Italy\\
$^{21}$ INAF-IASF Milano, Via Alfonso Corti 12, I-20133 Milano, Italy\\
$^{22}$ von Hoerner \& Sulger GmbH, Schlo{\ss}Platz 8, D-68723 Schwetzingen, Germany\\
$^{23}$ Aix-Marseille Univ, CNRS/IN2P3, CPPM, Marseille, France\\
$^{24}$ Institute for Computational Science, University of Zurich, Winterthurerstrasse 190, 8057 Zurich, Switzerland\\
$^{25}$ Institut de Physique Nucl\'eaire de Lyon, 4, rue Enrico Fermi, 69622, Villeurbanne cedex, France\\
$^{26}$ Universit\'e de Gen\`eve, D\'epartement de Physique Th\'eorique and Centre for Astroparticle Physics, 24 quai Ernest-Ansermet, CH-1211 Gen\`eve 4, Switzerland\\
$^{27}$ Institute of Theoretical Astrophysics, University of Oslo, P.O. Box 1029 Blindern, N-0315 Oslo, Norway\\
$^{28}$ Argelander-Institut f\"ur Astronomie, Universit\"at Bonn, Auf dem H\"ugel 71, 53121 Bonn, Germany\\
$^{29}$ Centre for Extragalactic Astronomy, Department of Physics, Durham University, South Road, Durham, DH1 3LE, UK\\
$^{30}$ Observatoire de Sauverny, Ecole Polytechnique F\'ed\'erale de Lau- sanne, CH-1290 Versoix, Switzerland\\
$^{31}$ Jet Propulsion Laboratory, California Institute of Technology, 4800 Oak Grove Drive, Pasadena, CA, 91109, USA\\
$^{32}$ Dipartimento di Fisica e Astronomia, Universit\'a di Bologna, Via Gobetti 93/2, I-40129 Bologna, Italy\\
$^{33}$ Institute of Space Sciences (IEEC-CSIC), c/Can Magrans s/n, 08193 Cerdanyola del Vall\'es, Barcelona, Spain\\
$^{34}$ Centro de Investigaciones Energ\'eticas, Medioambientales y Tecnol\'ogicas (CIEMAT), Avenida Complutense 40, 28040 Madrid, Spain\\
$^{35}$ Instituto de Astrof\'isica e Ci\^encias do Espa\c{c}o, Faculdade de Ci\^encias, Universidade de Lisboa, Tapada da Ajuda, PT-1349-018 Lisboa, Portugal\\
$^{36}$ Departamento de F\'isica, Faculdade de Ci\^encias, Universidade de Lisboa, Edif\'icio C8, Campo Grande, PT1749-016 Lisboa, Portugal\\
$^{37}$ Universidad Polit\'ecnica de Cartagena, Departamento de Electr\'onica y Tecnolog\'ia de Computadoras, 30202 Cartagena, Spain\\
$^{38}$ Infrared Processing and Analysis Center, California Institute of Technology, Pasadena, CA 91125, USA\\

\bibliographystyle{aa}
\bibliography{biblio}

\end{document}